\documentclass[letterpaper]{article}
\usepackage[T1]{fontenc}
\usepackage{geometry}
\usepackage{setspace}
\usepackage{amsmath}
\usepackage{amssymb}
\usepackage{bm}
\usepackage{booktabs}
\usepackage{graphicx}

\usepackage[articletitle=true, doi=true]{achemso}

\usepackage{authblk}
\author{Serguei V. Feskov*}
\author{Ivan F. Antipov}
\author{Anatoly I. Ivanov}
\affil{Volgograd State University, Volgograd, 400062, Russia}

\title{Excitonic and Charge-Transfer Contributions to Molecular Dimer Absorption: A Decomposition Approach Applied to a BPEA Dimer}

\date{*Email: serguei.feskov@volsu.ru}

\begin{document}

\maketitle

\begin{abstract}
    Electronic absorption spectra of multichromophoric systems are often governed by complex excited-state structures arising from excitonic and charge-transfer (CT) interactions between chromophores, while direct identification of the underlying electronic transitions is frequently hindered by strong vibronic and solvent-induced broadening. In this paper, we develop a theoretical framework for the analysis and decomposition of absorption spectra of molecular dimers with coupled Frenkel exciton (FE) and charge-transfer states, including solvent-induced stabilization of zwitterionic configurations within a unified adiabatic-state formalism. Our analysis reveals that exciton--CT mixing strongly reorganizes the electronic absorption profile and produces pronounced spectral broadening, while leaving the first spectral moment essentially unaffected. Numerical calculations show that the dominant mechanism of CT-induced broadening originates primarily from additional energetic splitting between spectral components rather than from broadening of the individual bands themselves. The electronic model is further extended to include coupling to high-frequency intramolecular vibrations and low-frequency environmental degrees of freedom, providing a practical framework for interpretation of realistic experimental spectra. The developed formalism is applied to the absorption spectrum of a covalently linked 9,10-bis(phenylethynyl)anthracene dimer in dichloromethane, where spectral decomposition reveals a predominantly excitonic low-energy doublet and higher-energy states with substantial CT character. The proposed approach offers a physically transparent framework for the analysis of complex absorption spectra in molecular aggregates and organic electronic materials with coupled excitonic and CT states.
\end{abstract}

\section*{Keywords}
bichromophoric systems; BPEA dimers; exciton and charge-transfer states; absorption spectra; vibronic coupling; solvent reorganization

\section{Introduction}

Organic molecular dimers composed of two identical chromophoric units, Ch$_\mathrm{A}$--Ch$_\mathrm{B}$, either covalently linked or weakly associated in solution, represent an important class of systems for investigating photoinduced processes in condensed phases.\cite{brixner_aem_17, terazono_jacs_15, sung_nc_15, wurthner_acr_16, zheng_jpcb_17, hong_jacs_22, HASHIMOTO2018, wega_pps_24, wega_jpcl_24, bressan_jpcc_25, reid_jpca_23, Wasielewski25b, PeryleneBisimide26} In such systems, electronic excitation is typically delocalized over the two chromophores and is strongly influenced by interchromophoric interactions as well as by coupling to the surrounding polar environment. As a result, molecular dimers provide a convenient framework for exploring excitation energy transfer and charge-transfer processes, as well as their interplay with intramolecular vibrational degrees of freedom and environmental fluctuations.\cite{roy_jacs_22, hernandez_jpcb_15, uranga_pccp_18, sissa_jpcb_10, kumar_jcp_21, valente_jcp_21, gultekin_jpca_24, IVANOVRev24, thakur_jpcb_24, yu_mol_25}

Beyond their role as model systems, molecular dimers are important components of optoelectronic and photocatalytic devices at the molecular scale.\cite{wasielewski_acr_09, Wasielewski2020, korovina_jcp_20, romero_ps_23} Particularly, bichromophoric architectures are widely employed in assemblies for light harvesting, charge separation, and energy conversion, where they often act as efficient photosensitizers. Their ability to support both excitonic delocalization and charge-transfer states makes them well suited for applications such as solar energy conversion and light-driven CO$_2$ transformation. \cite{wasielewski_jacs_04, kato_cr_15, mennucci_rmp_18, dunietz_jpcb_18, kellogg_fd_19, popp_arpc_21, imahori_amr_21, mukazhanova_jcp_21, machin2023, ghosh_acr_26}

While bichromophoric compounds offer significant advantages, their complex photophysical behavior also gives rise to a number of challenges. In particular, strong electronic coupling between Ch$_\mathrm{A}$ and Ch$_\mathrm{B}$ can open additional pathways for excited-state deactivation, thereby reducing the quantum yield of the target process. Two such pathways are commonly encountered: (i) excited-state symmetry breaking associated with intramolecular charge transfer (CT) between chromophores,\cite{Wasielewski_2017, Thompson18, Wasielewski21, Xia26} and (ii) formation of a quasi-stable excimer state exhibiting weak, red-shifted fluorescence relative to monomer emission. \cite{wasielewski_pccp_14, budyka_hec_17, bae2020, Wasielewski2020, chaudhuri_om_21, ma_jpcb_25, wang_cs_25} Both processes involve zwitterionic configurations, $\vert \mathrm{Ch}^+_\mathrm{A}\mathrm{Ch}^-_\mathrm{B} \rangle$ and $\vert \mathrm{Ch}^-_\mathrm{A}\mathrm{Ch}^+_\mathrm{B} \rangle$, as key intermediates.

The influence of CT states on the absorption spectra of molecular aggregates and multichromophoric systems has been extensively investigated through theoretical modeling, quantum-chemical calculations, and experimental spectroscopic studies.\cite{hestand_jcp_15, hestand_acr_17, hestand_cr_18, Wasielewski_2017, bialas_jpcc_19, feng_jcp_20, Wasielewski2020, canola_jcp_21, manrho_jcp_22, segalina_jctc_22, dai_mol_23, giannini_mt_24} Nevertheless, quantitative interpretation of experimental absorption spectra remains challenging because excitonic, charge-transfer, vibronic, and environmental effects contribute simultaneously to the observed spectral profiles. For this reason, the Kasha exciton model is often used as a simple framework for interpreting absorption spectra. In its original formulation, the model predicts the splitting of the monomer absorption band into an excitonic doublet, while zwitterionic states are assumed to possess negligible oscillator strength and are therefore not treated as direct contributors to the absorption profile.

In this paper, we explore the conditions under which intramolecular CT can contribute significantly not only to excited-state deactivation processes but also to the optical absorption of molecular dimers. In particular, we demonstrate that nonequilibrium polarization of the environment can promote the participation of zwitterionic states in the optical response of the dimer, enabling their contribution to the absorption spectrum alongside excitonic states. The interaction of the dimer with the fluctuating environment also leads to broadening of the absorption profile, with the magnitude of this broadening governed by the electronic coupling matrix elements associated with electron and hole transfer.

To describe photoinduced processes in bichromophoric compounds under nonequilibrium conditions, we employ the approach developed in Ref.~\citenum{Antipov22}. In addition to excitonic and CT interactions between the two chromophore units, this formulation explicitly accounts for the interaction of the dimer dipole moment with the nonequilibrium dielectric polarization of the environment. In contrast to approaches based on Holstein-type Hamiltonians,\cite{holstein_ap_59, barford_book_13, Spano22} it provides a theoretical framework for describing photoprocesses in dimers coupled to a low-frequency dielectric continuum. This feature makes the method suitable for the analysis of photoinduced processes in polar solvents.

The developed approach enables the theoretical analysis of absorption spectra of bichromophoric systems obtained in experiments. Spectroscopic characterization of organic chromophore dimers in polar solvents presents a significant challenge, as the underlying electronic structure is often obscured by pronounced spectral broadening.\cite{spano_acr_20} Optical excitation of $\pi$-conjugated chromophores is accompanied by a substantial redistribution of electron density, even in centrosymmetric systems with vanishing permanent dipole moments in the ground and excited states. As a result, the excitation process is strongly coupled to intramolecular vibrational degrees of freedom and to the dielectric polarization of the surrounding medium. The associated intramolecular and solvent reorganization leads to significant inhomogeneous broadening.\cite{bruschi_jcp_25} In bichromophoric systems, this effect often masks the fine structure associated with excitonic and zwitterionic states, making it difficult to resolve individual spectral contributions and to quantify the effects of interchromophoric interactions on the spectral lineshape. These considerations highlight the need for a theoretical framework capable of consistently accounting for environmental effects in the interpretation of experimental spectra.

\begin{figure}[ht]
  \centering
  \includegraphics[width=12 cm]{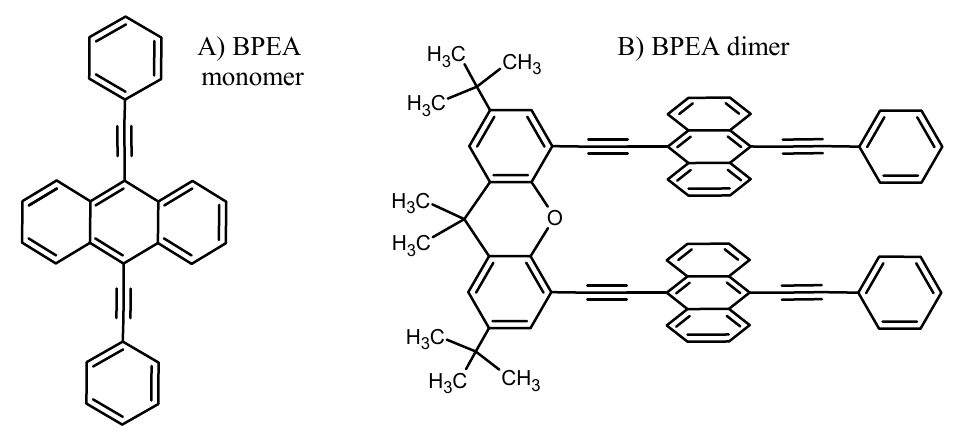}
  \caption{Molecular structures of (A) the monomer and (B) the covalently linked dimer studied in this work.}
  \label{fig:structures}
\end{figure}

As an illustration of the developed approach, we analyze the absorption spectrum of a molecular dimer consisting of two 9,10-bis(phenylethynyl)anthracene (BPEA) molecules covalently linked via a xanthene spacer (see Fig.~\ref{fig:structures}) in dichloromethane solution. This analysis provides insight into the mechanisms of inhomogeneous broadening arising from interchromophoric interactions. We examine the internal structure of the spectral profile and identify contributions associated with optical transitions to states of predominantly excitonic and zwitterionic character. The spectrum of the BPEA dimer is found to be dominated by the antisymmetric excitonic state, consistent with H-type aggregation. At the same time, charge-separated states, as well as the symmetric (lower) excitonic state, are also shown to make a significant contribution to the overall profile.

\section{Methods}
\label{sec:methods}


We consider a bichromophoric dimer in which each chromophore can undergo local electronic excitation as well as photoinduced charge separation, resulting in the formation of zwitterionic states. The diabatic basis functions of the dimer, corresponding to fixed electronic configurations of the $\mathrm{Ch}_\mathrm{A}$ and $\mathrm{Ch}_\mathrm{B}$ units, are defined as
\begin{equation} \label{phi_k_def}
    \vert \varphi_1 \rangle \equiv \vert \mathrm{Ch}^*_\mathrm{A}\mathrm{Ch}_\mathrm{B}\rangle, \quad
    \vert \varphi_2 \rangle \equiv \vert \mathrm{Ch}_\mathrm{A}\mathrm{Ch}^*_\mathrm{B}\rangle, \quad
    \vert \varphi_3 \rangle \equiv \vert \mathrm{Ch}^-_\mathrm{A}\mathrm{Ch}^+_\mathrm{B}\rangle, \quad
    \vert \varphi_4 \rangle \equiv \vert \mathrm{Ch}^+_\mathrm{A}\mathrm{Ch}^-_\mathrm{B}\rangle.
\end{equation}
Here, the asterisk ($^*$) denotes an electronically excited chromophore. Accordingly, $\vert \varphi_1 \rangle$ and $\vert \varphi_2 \rangle$ describe locally excited (LE) configurations, whereas $\vert \varphi_3 \rangle$ and $\vert \varphi_4 \rangle$ correspond to zwitterionic (charge-separated, CS) configurations. An arbitrary excited-state wavefunction $\vert \Psi_1 \rangle$ of the dimer is expressed as a superposition of these basis states,
\begin{equation} \label{Psi_1_decomp}
    \vert \Psi_1 \rangle = \sum_{k=1}^4 a_k \vert \varphi_k \rangle,
\end{equation}
where $a_k$ are complex coefficients satisfying the normalization condition $\sum_{k} |a_k|^2 = 1$.

\begin{figure}[ht]
   \centering
   \includegraphics[width=12 cm]{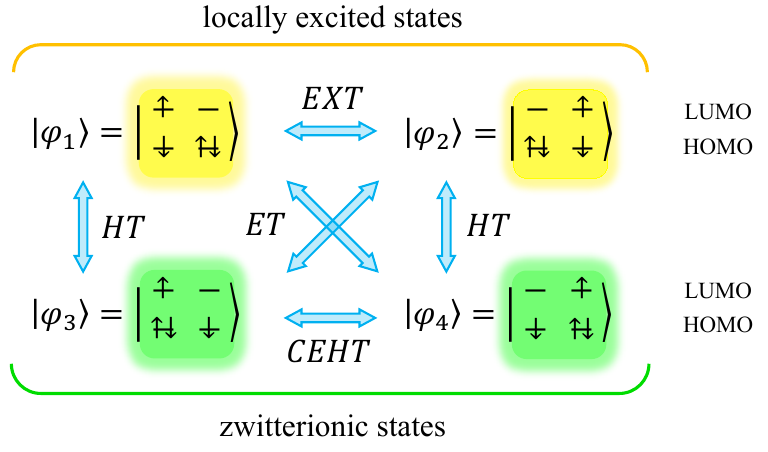}
   \caption{Diabatic excited-state configurations and the corresponding electronic couplings in a bichromophoric dimer. The abbreviations EXT, ET/HT and CEHT denote excitation energy transfer, electron/hole transfer, and concerted electron-hole transfer, respectively.}
   \label{fig:scheme}
\end{figure}

The electronic Hamiltonian of an isolated dimer in the $\vert \varphi_k \rangle$ basis is given by \cite{Spano22, Wasielewski2020, Antipov22}
\begin{equation}\label{H0_in_phi}
  \hat{H}_0 = \left(
    \begin{array}{cccc}
      0 & V_\mathrm{ext} & V_\mathrm{ht} & V_\mathrm{et} \\
      V_\mathrm{ext} & 0 & V_\mathrm{et} & V_\mathrm{ht} \\
      V_\mathrm{ht} & V_\mathrm{et} & \Delta E_\mathrm{cs} & V_\mathrm{ceht} \\[2pt]
      V_\mathrm{et} & V_\mathrm{ht} & V_\mathrm{ceht} & \Delta E_\mathrm{cs} 
    \end{array}
  \right),
\end{equation}
where the LE states are taken as the zero of energy, and $\Delta E_\mathrm{cs}$ denotes the energy of the (degenerate) CS states relative to the LE states. The parameter $V_\mathrm{ext}$ represents the electronic coupling between the two LE configurations, enabling excitation energy transfer between the chromophores, while $V_\mathrm{et}$ and $V_\mathrm{ht}$ describe electron and hole transfer arising from LUMO$_\mathrm{A}$--LUMO$_\mathrm{B}$ and HOMO$_\mathrm{A}$--HOMO$_\mathrm{B}$ interactions, respectively. The coupling between the two CS configurations $\vert \varphi_3 \rangle$ and $\vert \varphi_4 \rangle$ is given by $V_\mathrm{ceht}$. An illustration of the relevant electronic transitions in the bichromophoric dimer is shown in Fig.~\ref{fig:scheme}. This formulation is restricted to symmetric homodimers, but can be extended to heterodimeric systems by introducing non-degenerate diabatic state energies.

Nonequilibrium solvent effects are taken into account through the interaction of the dimer charge distribution with the orientational polarization of the solvent. The Hamiltonian of the photoexcited dimer in solution is given by \cite{Antipov22}
\begin{equation}\label{H_operator}
    \hat{H} = \hat{H}_0 - \lambda_\mathrm{or} D_\mathrm{m} \hat{D} + \frac{\lambda_\mathrm{or}}{2} D_\mathrm{m}^2 \hat{E},
\end{equation}
where $\lambda_\mathrm{or}$ is the interaction energy of a fully charge-separated state with the orientational solvent polarization, $D_\mathrm{m}$ is a dimensionless solvent coordinate describing the nonequilibrium polarization of the medium, and $\hat{D}$ is the charge-dissymmetry operator of the dimer. The definitions of $D_\mathrm{m}$ and $\hat{D}$ are given in Section 1 of the Supporting Information.

It is convenient to introduce the symmetry-adapted wavefunctions $\vert \psi_{1} \rangle$ and $\vert \psi_{2} \rangle$ representing delocalized Frenkel exciton states:
\begin{equation} \label{psi_k_basis}
    \vert \psi_{1,2} \rangle \equiv \frac{1}{\sqrt{2}} \big( \vert \varphi_1 \rangle \pm \vert \varphi_2 \rangle \big)
    = \frac{1}{\sqrt{2}} \big( \vert \mathrm{Ch}^*_\mathrm{A}\mathrm{Ch}_\mathrm{B} \rangle \pm \vert \mathrm{Ch}_\mathrm{A}\mathrm{Ch}^*_\mathrm{B} \rangle \big), 
    \qquad
    \vert \psi_{3,4} \rangle \equiv \vert \varphi_{3,4} \rangle.
\end{equation}
Here, the state $\vert \psi_{1} \rangle$ represents the symmetric, whereas $\vert \psi_{2} \rangle$ the antisymmetric, linear combination of the locally excited diabatic states $\vert \varphi_{1} \rangle$ and $\vert \varphi_{2} \rangle$. In the $\vert \psi_k \rangle$ basis, the Hamiltonian of the isolated dimer $\hat{H}_0$ takes the form
\begin{equation}\label{H0_in_psi}
  \hat{H}_0 = \left(
    \begin{array}{cccc}
      V_\mathrm{ext} & 0 & V_\Sigma & V_\Sigma \\
      0 & -V_\mathrm{ext} & -V_\Delta & V_\Delta \\
      V_\Sigma & -V_\Delta & \Delta E_\mathrm{cs} & V_\mathrm{ceht} \\
      V_\Sigma & V_\Delta & V_\mathrm{ceht} & \Delta E_\mathrm{cs} 
    \end{array}
  \right), 
\end{equation}
with
\begin{equation} \label{V_ds_def}
     V_\Delta \equiv \frac{1}{\sqrt{2}} \big( V_\mathrm{et} - V_\mathrm{ht} \big), \qquad
     V_\Sigma \equiv \frac{1}{\sqrt{2}} \big( V_\mathrm{et} + V_\mathrm{ht} \big). 
\end{equation}
This representation reveals the Davydov splitting between the two exciton states, $\Delta E_{12} = 2V_\mathrm{ext}$, as well as the symmetry-dependent coupling between the FE and CS configurations.\cite{hestand_jcp_15, hestand_cr_18, hoffmann_cp_00} In particular, the symmetric (upper) exciton state is coupled to the charge-separated configurations via the sum of the transfer integrals $V_\mathrm{et}$ and $V_\mathrm{ht}$, whereas the antisymmetric (lower) exciton state is coupled via their difference.

\subsection{Diabatic free energy surfaces}

The diagonal matrix elements of the Hamiltonian $\hat{H}$, evaluated as functions of the classical solvent coordinate $D_\mathrm{m}$, define the diabatic free energy surfaces (FESs),
\begin{equation}\label{FESs_def}
    G_k^{(\mathrm{d})}(D_\mathrm{m}) \equiv \langle \psi_k \vert \hat{H}(D_\mathrm{m}) \vert \psi_k \rangle,
\end{equation}
which describe the free energy of the bichromophoric dimer in a fixed electronic configuration $\vert \psi_k \rangle$ as a function of the nonequilibrium solvent polarization.

It is convenient to express the FES profiles in terms of the solvent reorganization energy associated with charge separation, $\lambda_\mathrm{cs}$. Since the interaction energy of a dipole with the orientational polarization of the solvent is twice the corresponding solvent reorganization energy, the two parameters are related by
\begin{equation*}
    \lambda_\mathrm{cs} = \frac{\lambda_\mathrm{or}}{2}.    
\end{equation*}
Using this definition and substituting Eq.~\eqref{H_operator} into Eq.~\eqref{FESs_def}, we obtain
\begin{equation} \label{diabatic_FESs}
   \begin{aligned}
     G_1^{(\mathrm{d})}(D_\mathrm{m}) &= \lambda_\mathrm{cs} D_\mathrm{m}^2 + V_\mathrm{ext}, \\
     G_2^{(\mathrm{d})}(D_\mathrm{m}) &= \lambda_\mathrm{cs} D_\mathrm{m}^2 - V_\mathrm{ext}, \\
     G_3^{(\mathrm{d})}(D_\mathrm{m}) &= \lambda_\mathrm{cs} D_\mathrm{m}^2 - 2\lambda_\mathrm{cs} D_\mathrm{m} + \Delta E_\mathrm{cs} = \lambda_\mathrm{cs} \left( D_\mathrm{m} - 1 \right)^2 - \lambda_\mathrm{cs} + \Delta E_\mathrm{cs}, \\
     G_4^{(\mathrm{d})}(D_\mathrm{m}) &= \lambda_\mathrm{cs} D_\mathrm{m}^2 + 2\lambda_\mathrm{cs} D_\mathrm{m} + \Delta E_\mathrm{cs} = \lambda_\mathrm{cs} \left( D_\mathrm{m} + 1 \right)^2 - \lambda_\mathrm{cs} + \Delta E_\mathrm{cs}.
   \end{aligned}
\end{equation}
All diabatic FESs $G_k^{(\mathrm{d})}(D_\mathrm{m})$ are parabolic functions of $D_\mathrm{m}$ with curvature determined by the solvent reorganization energy $\lambda_\mathrm{cs}$. This parabolic structure arises directly from the linear solute--solvent coupling assumed in Eq.~\eqref{H_operator}. Similar diabatic free-energy surfaces appear in the Marcus theory of electron transfer, where the solvent is described in terms of a collective polarization coordinate.\cite{Marcus56}

\begin{figure}[ht]
   \centering
   \includegraphics[width=12 cm]{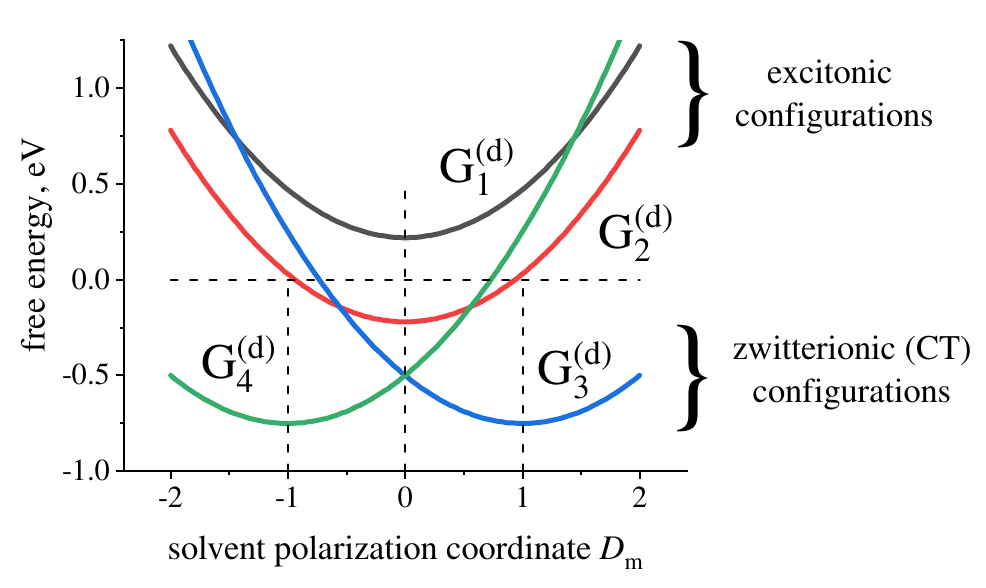}
   \caption{Diabatic free energy surfaces $G_k^{(\mathrm{d})}(D_\mathrm{m})$ corresponding to the lowest excited states of the bichromophoric dimer in a polar solvent, plotted as functions of the solvent coordinate $D_\mathrm{m}$. The parabolic profiles reflect linear solute--solvent coupling. The surfaces are calculated using Eq.~\eqref{diabatic_FESs} with $V_\mathrm{ext} = 0.22$~eV, $\lambda_\mathrm{cs} = 0.25$~eV, and $\Delta E_\mathrm{cs} = -0.5$~eV.}
   \label{fig:dia_fess}
\end{figure}

The minima of the diabatic FESs $G_k^{(\mathrm{d})}$ define the equilibrium solvent coordinates $\widetilde{D}_\mathrm{m}^{(k)}$ for the corresponding electronic states. From Eq.~\eqref{diabatic_FESs}, one obtains $\widetilde{D}_\mathrm{m}^{(1)} = \widetilde{D}_\mathrm{m}^{(2)} = 0$ for the excitonic states and $\widetilde{D}_\mathrm{m}^{(3)} = +1$, $\widetilde{D}_\mathrm{m}^{(4)} = -1$ for the CS states. Representative diabatic surfaces are shown in Fig.~\ref{fig:dia_fess}.

The diabatic framework enables description of chemical and spectroscopic dynamics of photoexcited bichromophoric systems in terms of wavepacket motion and nonadiabatic electronic transitions at surface crossings. The validity of this description requires
\begin{equation} \label{diabatic_limits}
    V_\Sigma, V_\Delta, V_\mathrm{ceht} \ll k_\mathrm{B}T,
\end{equation}
which ensures that charge transfer occurs in the nonadiabatic regime, such that transitions can be described in terms of weak coupling between diabatic states.

Within this representation, charge transfer between Ch$_\mathrm{A}$ and Ch$_\mathrm{B}$ is described as a reversible quantum transition between excitonic and charge-separated surfaces in the vicinity of their intersection points (Fig.~\ref{fig:dia_fess}). The rate of such transitions is determined by the corresponding off-diagonal matrix elements of $\hat{H}_0$. From Eqs.~\eqref{diabatic_FESs} and \eqref{H0_in_psi}, the coordinates of the surface crossings and the associated electronic coupling elements are obtained as
\begin{equation} \label{cross_points}
    \begin{aligned}
        \vert \psi_1 \rangle \leftrightarrow \vert \psi_3 \rangle:& \quad \check{D}_\mathrm{m}^{(1,3)} =  -(\Delta E_\mathrm{cs} - V_\mathrm{ext}) / 2\lambda_\mathrm{cs}, \quad V = V_\Sigma, \\
        \vert \psi_1 \rangle \leftrightarrow \vert \psi_4 \rangle:& \quad \check{D}_\mathrm{m}^{(1,4)} =  +(\Delta E_\mathrm{cs} - V_\mathrm{ext}) / 2\lambda_\mathrm{cs}, \quad V = V_\Sigma, \\
        \vert \psi_2 \rangle \leftrightarrow \vert \psi_3 \rangle:& \quad \check{D}_\mathrm{m}^{(2,3)} =  -(\Delta E_\mathrm{cs} + V_\mathrm{ext}) / 2\lambda_\mathrm{cs}, \quad V = -V_\Delta, \\
        \vert \psi_2 \rangle \leftrightarrow \vert \psi_4 \rangle:& \quad \check{D}_\mathrm{m}^{(2,4)} =  +(\Delta E_\mathrm{cs} + V_\mathrm{ext}) / 2\lambda_\mathrm{cs}, \quad V = V_\Delta, \\
        \vert \psi_3 \rangle \leftrightarrow \vert \psi_4 \rangle:& \quad \check{D}_\mathrm{m}^{(3,4)} =  0, \quad V = V_\mathrm{ceht}.
    \end{aligned}    
\end{equation}
The first column specifies the electronic states involved in the transition, the second gives the corresponding crossing coordinate $\check{D}_\mathrm{m}$, and the third lists the associated electronic coupling element.

\subsection{Adiabatic Representation under Strong Charge-Transfer Coupling}

We now consider the regime of strong interchromophoric coupling, in which at least one of the matrix elements responsible for intramolecular charge transfer ($V_\Sigma$, $V_\Delta$, or $V_\mathrm{ceht}$) is comparable to or exceeds $k_\mathrm{B}T$. In this regime, the diabatic representation is no longer adequate in the vicinity of the crossing points $\check{D}_\mathrm{m}^{(k,k')}$, where electronic coupling induces substantial mixing of diabatic states. Consequently, the crossings between diabatic FESs are replaced by avoided crossings in the adiabatic representation.

To describe this regime, we introduce the adiabatic FESs $G_k^{(\mathrm{a})}$ as the eigenvalues of the Hamiltonian $\hat{H}$, obtained from the parameter-dependent eigenvalue problem
\begin{equation}\label{eigensolution}
    \left( \hat{H}_0 - \lambda_\mathrm{or} D_\mathrm{m} \hat{D} + \frac{\lambda_\mathrm{or}}{2} D_\mathrm{m}^2 \hat{E} \right) \vert \Psi_k^{(\mathrm{a})}(D_\mathrm{m}) \rangle = G_k^{(\mathrm{a})}(D_\mathrm{m}) \vert \Psi_k^{(\mathrm{a})}(D_\mathrm{m}) \rangle.  
\end{equation}
Here, $\vert \Psi_k^{(\mathrm{a})} \rangle$ denotes the $k$-th adiabatic eigenstate associated with the eigenvalue $G_k^{(\mathrm{a})}$. The solvent coordinate $D_\mathrm{m}$ enters Eq.~\eqref{eigensolution} as an external parameter, such that both $\vert \Psi_k^{(\mathrm{a})}(D_\mathrm{m}) \rangle$ and $G_k^{(\mathrm{a})}(D_\mathrm{m})$ are parametric functions of $D_\mathrm{m}$. 

Since the eigenvalues of $\hat{H}$ are invariant under a change of basis, the adiabatic FESs can be obtained by diagonalizing the full Hamiltonian in either of the representations, Eqs.~\eqref{H0_in_phi} or \eqref{H0_in_psi}. In the limit of vanishing CT-couplings ($V_\mathrm{et}, V_\mathrm{ht}, V_\mathrm{ceht} \to 0$) the adiabatic FESs $G_k^{(\mathrm{a})}$ reduce to the corresponding diabatic surfaces $G_k^{(\mathrm{d})}$ defined in Eq.~\eqref{diabatic_FESs}.

\begin{figure}[ht]
   \centering
   \includegraphics[width=16.0 cm]{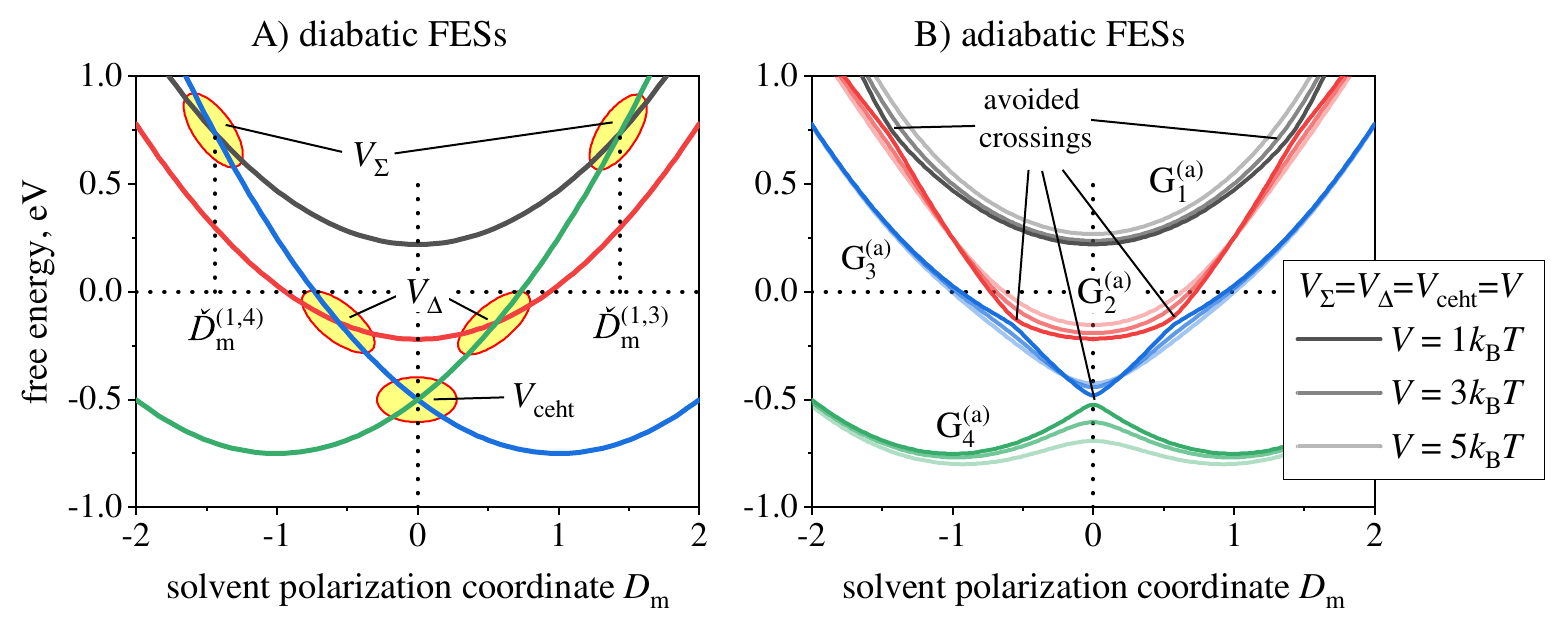}
   \caption{Effect of charge-transfer couplings ($V_\Sigma$, $V_\Delta$, and $V_\mathrm{ceht}$) on the adiabatic free energy surfaces of the lowest excited states of the dimer. (A) Diabatic FESs with crossing regions (highlighted) and the corresponding coupling parameters governing CT transitions. (B) Adiabatic FESs $G_k^{(\mathrm{a})}$ obtained from Eq.~\eqref{eigensolution} for different coupling strengths (as indicated). Other parameters are the same as in Fig.~\ref{fig:dia_fess}.}
   \label{fig:adia_fess}
\end{figure}

In general, the eigenvalue problem in Eq.~\eqref{eigensolution} does not admit a closed-form analytical solution and is therefore solved numerically without introducing additional physical approximations. The resulting numerical solutions are used below to analyze the effect of interchromophoric CT-couplings ($V_\Sigma$, $V_\Delta$, $V_\mathrm{ceht}$) on the adiabatic FES profiles. Figure~\ref{fig:adia_fess}B shows the  $G_k^{(\mathrm{a})}(D_\mathrm{m})$ surfaces calculated for different CT coupling strengths. For clarity, all three coupling parameters were set equal and varied simultaneously from $k_\mathrm{B}T$ to $5k_\mathrm{B}T$.

The resulting profiles show characteristic avoided crossings in the vicinity of the intersection points of the diabatic FESs. The magnitude of each splitting is determined by the corresponding off-diagonal coupling element. For example, the splitting between $G_1^{(\mathrm{a})}$ and $G_3^{(\mathrm{a})}$ near $\check{D}_\mathrm{m}^{(1,3)}$ and $\check{D}_\mathrm{m}^{(1,4)}$ is governed by $V_\Sigma$, whereas that between $G_2^{(\mathrm{a})}$ and $G_3^{(\mathrm{a})}$ near $\check{D}_\mathrm{m}^{(2,3)}$ and $\check{D}_\mathrm{m}^{(2,4)}$ is controlled by $V_\Delta$. The splitting between $G_3^{(\mathrm{a})}$ and $G_4^{(\mathrm{a})}$ is determined by $V_\mathrm{ceht}$. These relationships are schematically summarized in Fig.~\ref{fig:adia_fess}A. The observed level repulsion is a standard consequence of electronic state mixing; in the limit of weak electronic coupling $V$, the energy gap at the avoided crossing is $\Delta E \approx 2|V|$.\cite{cohen_book_92}

The set of adiabatic surfaces $G_k^{(\mathrm{a})}(D_\mathrm{m})$ offers a convenient framework for describing photochemical processes in compact dimers with strong interchromophoric interactions, $V_\Sigma, V_\Delta, V_\mathrm{ceht} \gtrsim k_\mathrm{B}T$. In such systems, the lowest excited states are separated by large energy gaps $\Delta E \gtrsim k_\mathrm{B}T$, which reduces the probability of nonadiabatic transitions between the adiabatic states in the vicinity of avoided crossings. As a result, the dynamics is largely governed by wave packet motion on the adiabatic FESs, with transitions to lower-lying states occurring predominantly via vertical radiative or nonradiative processes.

\subsection{Optical Activity of Adiabatic States}

The absorption spectra of multichromophoric systems are determined not only by their energetic structure but also by the transition dipole moments (TDMs) associated with the corresponding excited states.\cite{Antipov22} In strongly coupled dimers, these moments are governed by the symmetry and electronic composition of the adiabatic states.

Consider the optical transition from the ground state $\vert \Psi_0 \rangle$ to an arbitrary excited state $\vert \Psi_1 \rangle = \left( a_1, a_2, a_3, a_4 \right)^\mathrm{T}$ expressed in the $\vert \varphi_k \rangle$ basis. The corresponding transition dipole moment is
\begin{equation}\label{vector_mu_opt1}
    \bm{\mu}_\mathrm{opt} = \langle \Psi_0 \vert \hat{\bm{\mu}}_\mathrm{opt} \vert \Psi_1 \rangle
    = \sum\limits_k a_k \bm{\mu}_{0k},
\end{equation}
where $\bm{\mu}_{0k} \equiv \langle \Psi_0 \vert \hat{\bm{\mu}}_\mathrm{opt} \vert \varphi_k \rangle$ denotes the transition dipole moment between $\vert \Psi_0 \rangle$ and $\vert \varphi_k \rangle$.

We assume that direct optical transitions to the zwitterionic states $\vert \mathrm{Ch}^-_\mathrm{A} \mathrm{Ch}^+_\mathrm{B} \rangle$ and $\vert \mathrm{Ch}^+_\mathrm{A} \mathrm{Ch}^-_\mathrm{B} \rangle$ are negligible, so that $\bm{\mu}_{03} = \bm{\mu}_{04} = 0$. For identical chromophores we have $\mu_{01} = \mu_{02} \equiv \mu_0$, where $\mu_0$ is the transition dipole moment of an isolated chromophore. If $\theta$ denotes the angle between the local TDMs $\bm{\mu}_{01}$ and $\bm{\mu}_{02}$, the squared transition dipole moment becomes
\begin{equation}\label{sqr_mu_opt2}
    \mu^2_\mathrm{opt} / \mu_0^2 =
    \vert a_1 \vert^2 + \vert a_2 \vert^2
    + \cos\theta \left( a_1 a_2^* + a_1^* a_2 \right).
\end{equation}
The derivation of Eq.~\eqref{sqr_mu_opt2}, including its particular forms for parallel, perpendicular, and antiparallel TDM orientations, is given in Section 2 of the Supporting Information.

\begin{figure}[ht]
   \centering
   \includegraphics[width=14 cm]{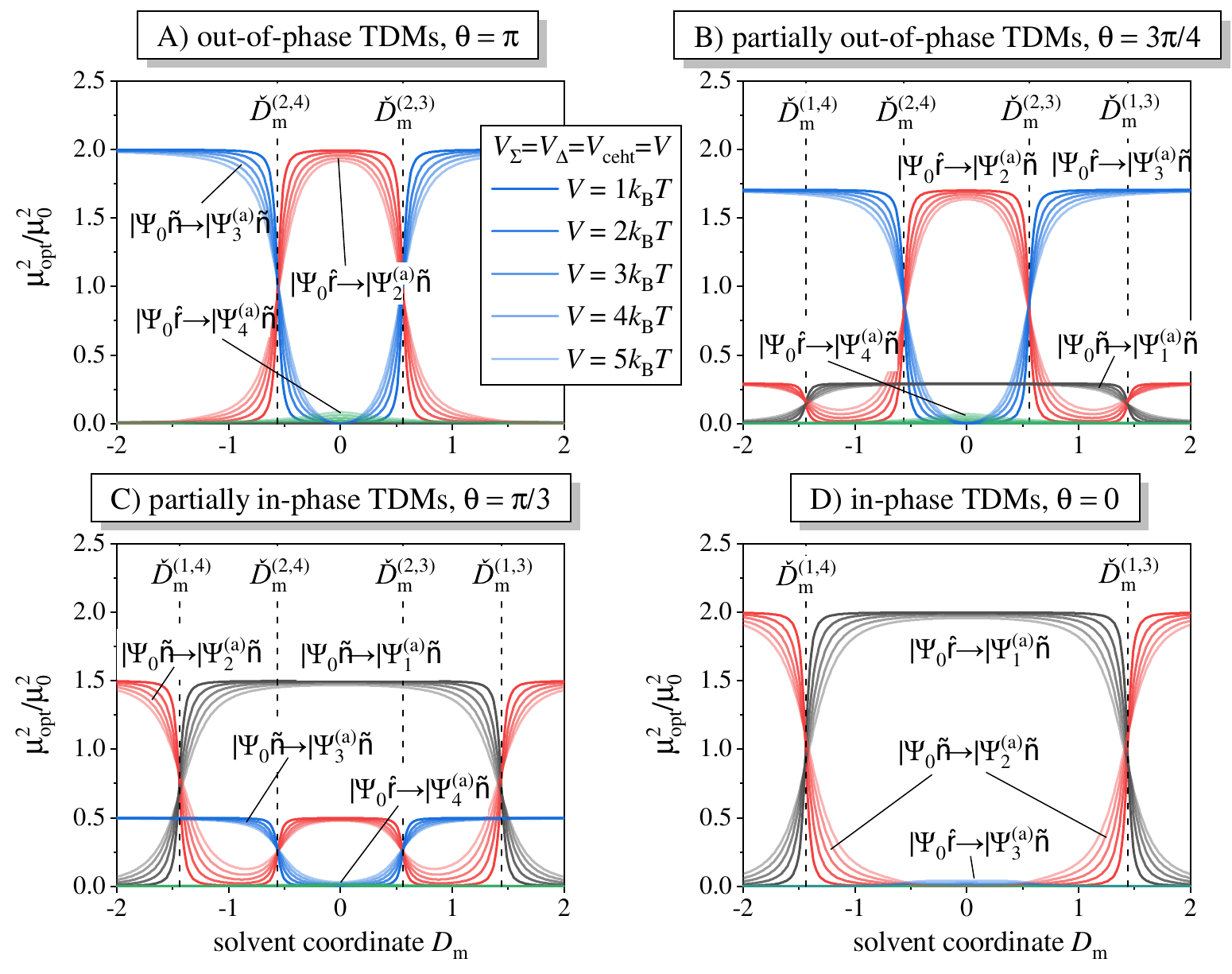}
   \caption{Normalized squared transition dipole moments $\mu^2_\mathrm{opt}/\mu_0^2$ for optical transitions from $\vert \Psi_0 \rangle$ to the lowest adiabatic states $\vert \Psi_k^{(\mathrm{a})} \rangle$ ($k = 1$--4), plotted as functions of the solvent coordinate $D_\mathrm{m}$. The results are obtained from Eq.~\eqref{sqr_mu_opt2} using eigenstates calculated from Eq.~\eqref{eigensolution}. Panels A--D correspond to different angles between the local TDMs $\bm{\mu}_{01}$ and $\bm{\mu}_{02}$: $\theta = \pi$, $3\pi/4$, $\pi/3$, and $0$ (from out-of-phase to in-phase alignment). Other parameters are the same as in Fig.~\ref{fig:dia_fess}. Symbols $\check{D}^{(k,k')}_\mathrm{m}$ mark the quasi-crossings of the adiabatic FESs $G_k^{(\mathrm{a})}(D_\mathrm{m})$ and $G_{k'}^{(\mathrm{a})}(D_\mathrm{m})$, as defined in Eq.~\eqref{cross_points}.}
\label{fig:tdm}
\end{figure}

The expression above shows that the optical activity of an adiabatic state depends on both its electronic composition and the relative orientation of the local transition dipole moments. Because the expansion coefficients $a_k$ of the adiabatic states vary with the solvent coordinate $D_\mathrm{m}$, the corresponding optical TDMs can also vary along this coordinate. This dependence is illustrated in Fig.~\ref{fig:tdm}, which shows the normalized squared TDMs $\mu^2_\mathrm{opt}/\mu_0^2$ for transitions from the ground state $\vert \Psi_0 \rangle$ to the four lowest adiabatic states $\vert \Psi_k^{(\mathrm{a})} \rangle$ ($k=1$--4). The calculations are shown for several values of the angle $\theta$ (panels A--D) and for coupling strengths $V = V_\Sigma = V_\Delta = V_\mathrm{ceht}$ ranging from $1\,k_\mathrm{B}T$ to $5\,k_\mathrm{B}T$. The curves corresponding to different $\vert \Psi_k^{(\mathrm{a})} \rangle$ are color-coded to match the adiabatic FESs shown in Fig.~\ref{fig:adia_fess}B.

The plots in Fig.~\ref{fig:tdm} illustrate the evolution of the optical activity of the adiabatic excited states $\vert \Psi_k^{(\mathrm{a})} \rangle$ resulting from solvent-induced mixing of electronic configurations. The most pronounced changes in $\mu^2_\mathrm{opt}$ occur near the quasi-crossing points $\check{D}_\mathrm{m}^{(k,k')}$ indicated in the figure. These changes are strongest for small electronic couplings $V$, whereas increasing $V$ leads to a smoother dependence on $D_\mathrm{m}$. This behavior reflects the evolution of the electronic composition of the adiabatic states, with FE-dominated states transforming into CT-dominated states and vice versa; the transformation becomes increasingly gradual as $V$ increases (see also Fig.~\ref{fig:adia_fess}B).

A comparison of panels A--D in Fig.~\ref{fig:tdm} further demonstrates the influence of $\theta$ on the optical activity of the adiabatic states. For $\theta = \pi$, corresponding to antiparallel local TDMs, the main optical intensity is associated with the odd exciton state $\vert \psi_2 \rangle$. Consequently, the brightest transitions occur to adiabatic states with substantial $\vert \psi_2 \rangle$ character, namely $\vert \Psi_2^{(\mathrm{a})} \rangle$ and $\vert \Psi_3^{(\mathrm{a})} \rangle$ within the corresponding ranges of $D_\mathrm{m}$, as follows from the FES profiles in Fig.~\ref{fig:adia_fess}A.

As $\theta$ decreases from $\pi$ (panel B), the optical activity of $\vert \Psi_2^{(\mathrm{a})} \rangle$ and $\vert \Psi_3^{(\mathrm{a})} \rangle$ decreases, whereas $\vert \Psi_1^{(\mathrm{a})} \rangle$ becomes increasingly bright. Its intensity reaches a maximum in the range $\check{D}_\mathrm{m}^{(1,4)} \lesssim D_\mathrm{m} \lesssim \check{D}_\mathrm{m}^{(1,3)}$, where $\vert \Psi_1^{(\mathrm{a})} \rangle$ is predominantly described by the even exciton state $\vert \psi_1 \rangle$. Thus, variation of $\theta$ modifies the interference between the two excitation pathways and redistributes oscillator strength among the adiabatic states. Deviations from $\theta=\pi$ consequently give rise to an additional absorption contribution associated with the upper adiabatic state. Its intensity increases as $\theta$ decreases and becomes dominant for $\theta < \pi/2$ (panels C and D).

Overall, the optical activity of the adiabatic states is governed by the interplay of electronic symmetry, the relative orientation of the local TDMs, and solvent-induced mixing of FE and CT configurations. The state-resolved analysis of $\mu^2_\mathrm{opt}$ provides insight into the redistribution of transition strength, while experimental spectra reflect the combined contributions of all optically active transitions. In the following section, the calculated energies and transition dipole moments are used to construct the absorption spectrum of the dimer and relate its spectral features to the underlying electronic structure.

\section{Results and Discussion}

\subsection{Absorption Spectra: CT-Driven Electronic Structure and Vibronic Broadening}\label{sec:spectra}

Within the adiabatic framework developed in Section~\ref{sec:methods}, the absorption spectrum of the dimer is determined by (i) the $D_\mathrm{m}$-dependent energy gaps between the ground-state FES $G_0(D_\mathrm{m})$ and the excited-state surfaces $G_k^{(\mathrm{a})}(D_\mathrm{m})$ ($k = 1,\dots,4$), and (ii) the corresponding transition strengths, defined by the squared transition dipole moments of the adiabatic states.

Similar to Eqs.~\eqref{diabatic_FESs}, the ground-state FES in the linear-response approximation is a parabolic function of $D_\mathrm{m}$, with curvature determined by $\lambda_\mathrm{cs}$,
\begin{equation}\label{G0(Dm)}
    G_0(D_\mathrm{m}) = \lambda_\mathrm{cs} D_\mathrm{m}^2 - E_\mathrm{ex}.
\end{equation}
Here, $E_\mathrm{ex}$ is the excitation energy of an isolated chromophore.

Prior to photoexcitation, the system is in thermal equilibrium, and the distribution of the solvent coordinate $D_\mathrm{m}$ is given by the Boltzmann law,
\begin{equation*}
    \rho_0(D_\mathrm{m}) = A_0 \exp\left[-\frac{G_0(D_\mathrm{m})}{k_\mathrm{B}T}\right],
\end{equation*}
where $A_0$ is the normalization constant. For the quadratic potential \eqref{G0(Dm)}, this distribution reduces to a Gaussian function centered at $D_\mathrm{m} = 0$, with a standard deviation $\sigma_0 = \sqrt{k_\mathrm{B}T / 2\lambda_\mathrm{cs}}$:
\begin{equation}\label{rho_0_boltzmann}
    \rho_0(D_\mathrm{m}) = \frac{1}{\sqrt{2\pi \sigma_0^2}} 
    \exp\left( - \frac{D_\mathrm{m}^2}{2\sigma_0^2} \right).
\end{equation}

The total absorption profile of the dimer is obtained as a sum of contributions corresponding to optical transitions from $\vert \Psi_0 \rangle$ to the adiabatic excited states $\vert \Psi_k^{(\mathrm{a})} \rangle$,
\begin{equation}\label{spectrum_def}
    S_0(\hbar\omega) = \sum_{k=1}^{4} \int\limits_{-\infty}^{+\infty} 
    \rho_0(D_\mathrm{m})\, \frac{\mu_{\mathrm{opt},k}^2(D_\mathrm{m})}{2\mu_0^2}\, 
    \delta\!\left( \Delta G_{0k}(D_\mathrm{m}) - \hbar\omega \right) \, dD_\mathrm{m},
\end{equation}
where $\mu_{\mathrm{opt},k}(D_\mathrm{m})$ is the TDM associated with the transition $\vert \Psi_0 \rangle \to \vert \Psi_k^{(\mathrm{a})} \rangle$, and $\Delta G_{0k}(D_\mathrm{m}) \equiv G_k^{(\mathrm{a})}(D_\mathrm{m}) - G_0(D_\mathrm{m})$ denotes the corresponding free energy gap. The factor $1/(2\mu_0^2)$ provides normalization with respect to the total oscillator strength of an isolated dimer consisting of two identical chromophores, each characterized by the transition dipole moment $\mu_0$. The Dirac delta function enforces energy conservation, selecting solvent configurations for which the transition energy matches the photon energy $\hbar\omega$.

Although Eq.~\eqref{spectrum_def} generally requires numerical evaluation, an analytical solution can be obtained in the limiting case $V_\mathrm{et} = V_\mathrm{ht} = V_\mathrm{ceht} = 0$, i.e., in the absence of CT interactions. In this limit, the eigenvalues of $\hat{H}$ reduce to the diabatic FES given by Eq.~\eqref{diabatic_FESs}, and the energy gaps $\Delta G_{01}$ and $\Delta G_{02}$, as well as the corresponding transition dipole moments $\mu_{01}$ and $\mu_{02}$, become independent of $D_\mathrm{m}$. One then obtains
\begin{equation} \label{analytic_1}
    \begin{aligned}
        \Delta G_{01} &= E_\mathrm{ex} + V_\mathrm{ext}, \qquad \mu_{01}^2/\mu_0^2 = (1 + \cos\theta), \\
        \Delta G_{02} &= E_\mathrm{ex} - V_\mathrm{ext}, \qquad \mu_{02}^2/\mu_0^2 = (1 - \cos\theta).
    \end{aligned}    
\end{equation}
Since only the first two terms in Eq.~\eqref{spectrum_def} contribute, the spectrum simplifies to
\begin{equation} \label{spectrum_2delta}
    S_0^\mathrm{(FE)}(\hbar\omega) = \frac{1 + \cos\theta}{2}\,\delta(E_\mathrm{ex} + V_\mathrm{ext} - \hbar\omega) 
    + \frac{1 - \cos\theta}{2}\,\delta(E_\mathrm{ex} - V_\mathrm{ext} - \hbar\omega).
\end{equation}

This expression reproduces the main result of the Kasha model,\cite{Kasha1965} in which the dimer absorption spectrum consists of two narrow lines corresponding to optical transitions to the FE states $\vert \psi_1 \rangle$ and $\vert \psi_2 \rangle$, split by $\pm V_\mathrm{ext}$ with respect to $\hbar\omega_0 = E_\mathrm{ex}$. Their intensities depend on the relative orientation of the monomer transition dipoles. For $\theta = 0$ (in-phase local TDMs), only the high-frequency component $\hbar\omega_0^{(\mathrm{high})} = E_\mathrm{ex} + V_\mathrm{ext}$ is optically allowed, whereas for $\theta = \pi$ (out-of-phase TDMs) only the low-frequency component $\hbar\omega_0^{(\mathrm{low})} = E_\mathrm{ex} - V_\mathrm{ext}$ is present.\cite{Kasha1965, hestand_cr_18}

The agreement between Eq.~\eqref{spectrum_2delta} and the Kasha model in the pure excitonic limit shows that the absorption profile retains its form even in the presence of solvent degrees of freedom. In particular, $S_0^\mathrm{(FE)}$ is independent of $\lambda_\mathrm{cs}$, indicating that the spectrum is insensitive to solvent polarization in the absence of CT couplings. Since the excited states do not involve charge separation in this regime, they couple only weakly to the solvent environment.

Using the analytical result in Eq.~\eqref{spectrum_2delta} as a reference, we now examine the effect of CT interactions on the absorption profile $S_0(\hbar\omega)$. As a starting point, we consider the case $\theta = \pi$, $V_\mathrm{et} = - V_\mathrm{ht}$, and $V_\mathrm{ceht} = 0$. In this configuration, the upper exciton state $\vert \psi_1 \rangle$ is optically dark and uncoupled from the CT states ($\mu_{01} = 0$, $V_\Sigma = 0$), effectively reducing the complexity of the spectral structure. Nevertheless, this particular case captures the essential mechanisms of CT-induced spectral modifications. The resulting physical picture can be readily generalized to more complex situations, including arbitrary orientations and relative phases of the local TDMs, as well as finite coupling between the upper exciton and zwitterionic states.

\begin{figure}[htbp]
   \centering
   \includegraphics[width=13.0 cm]{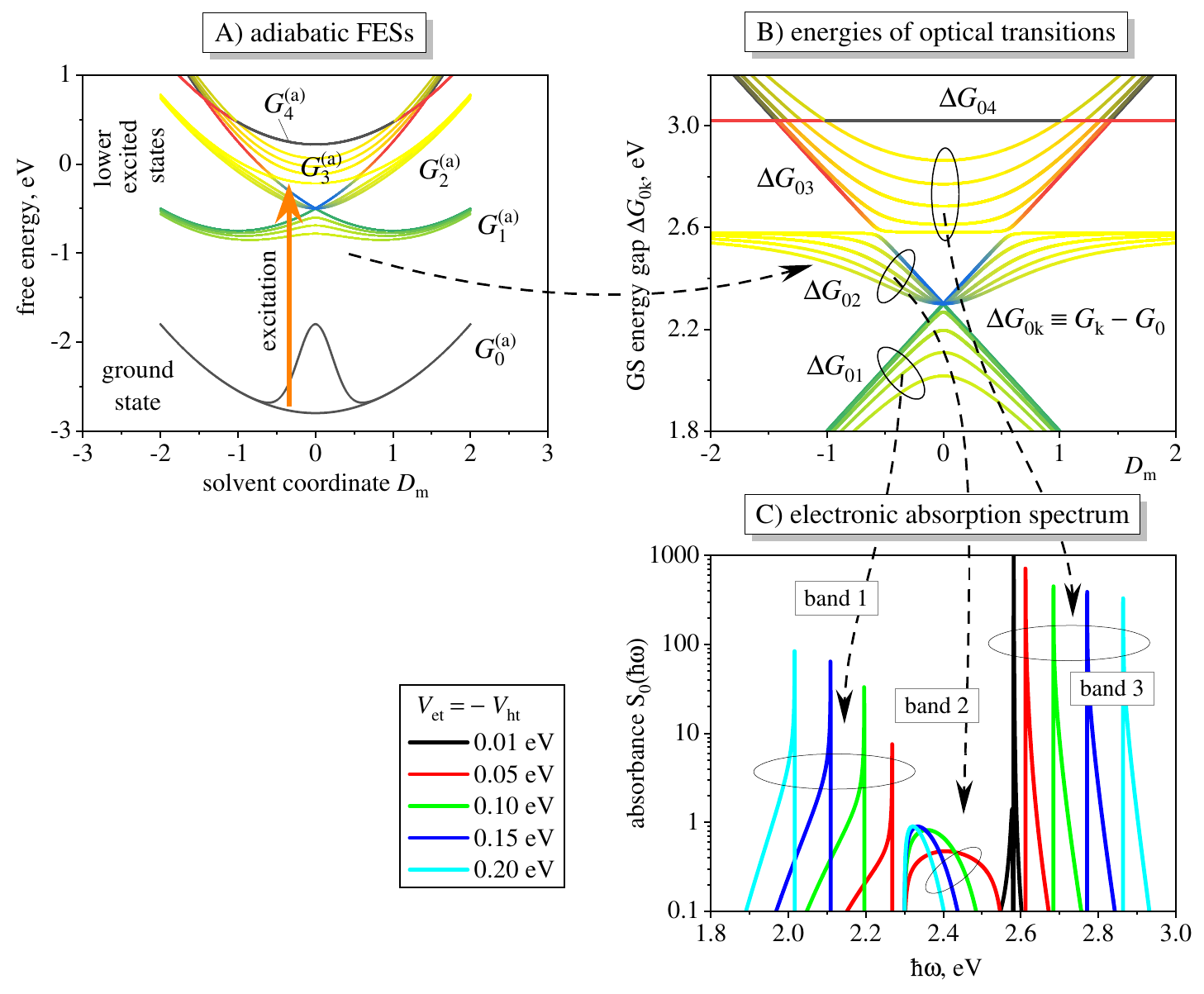}
   \caption{Effect of the charge-transfer coupling on the absorption spectrum of the dimer. 
   \textbf{A}: Adiabatic free energy surfaces $G_k^{(\mathrm{a})}(D_\mathrm{m})$ for the ground and low-lying excited states at different values of $V_\mathrm{et} = -V_\mathrm{ht}$. Yellow shading along the curves indicates the relative transition dipole strength $\mu^2_{\mathrm{opt},k}/\mu_0^2$. 
   \textbf{B}: Corresponding optical transition energies $\Delta G_{0k}(D_\mathrm{m}) = G_k^{(\mathrm{a})}(D_\mathrm{m}) - G_0(D_\mathrm{m})$, with the same color encoding of transition strengths. 
   \textbf{C}: CT-driven absorption spectra $S_0(\hbar\omega)$ calculated from Eq.~\eqref{spectrum_def} for the same parameter values; bands 1–3 correspond to transitions to $\vert \Psi_1^{(\mathrm{a})} \rangle$, $\vert \Psi_2^{(\mathrm{a})} \rangle$, and $\vert \Psi_3^{(\mathrm{a})} \rangle$, respectively. 
   Parameters: $\theta = \pi$, $\lambda_\mathrm{cs} = 0.25$~eV, $\Delta E_\mathrm{cs} = -0.5$~eV, $V_\mathrm{ext} = 0.22$~eV, $V_\mathrm{ceht} = 0$, and $V_\mathrm{et} = -V_\mathrm{ht}$ as indicated.}
   \label{fig:spectra}
\end{figure}

We now analyze the absorption spectra obtained numerically for this particular case: $\theta = \pi$, $V_\mathrm{ceht} = V_\Sigma = 0$. Figure~\ref{fig:spectra} summarizes the resulting spectra $S_0(\hbar\omega)$ for coupling strengths $V_\mathrm{et} = -V_\mathrm{ht}$ ranging from 0.01 to 0.2~eV, together with the associated energetic structure and optical activity of the adiabatic states. The figure illustrates successive stages in the formation of the absorption spectrum. As seen from panels A and B, increasing CT coupling leads to pronounced distortions of the adiabatic FESs and the corresponding transition energies $\Delta G_{0k}(D_\mathrm{m})$, accompanied by a redistribution of oscillator strength along the coordinate $D_\mathrm{m}$. These effects are most significant in the vicinity of quasi-crossing regions, where mixing between excitonic and charge-transfer configurations strongly modifies both the energies and intensities of optical transitions.

The resulting spectral changes are shown in panel C. With increasing $V_\mathrm{et}$, the initially narrow excitonic band (black curve) evolves into a broader and more structured profile containing contributions from several adiabatic states. The spectral evolution originates from the combined effect of exciton--CT mixing and the distribution of transition energies over thermally populated solvent configurations, resulting in pronounced inhomogeneous broadening of the spectrum.

As shown in Fig.~\ref{fig:spectra}C, increasing the coupling between the lower FE and CT states leads to a pronounced blue shift and broadening of the main excitonic absorption band (band 3 in the figure). Simultaneously, two additional low-energy bands (bands 1 and 2) emerge and progressively gain intensity with increasing $V_\mathrm{et} = -V_\mathrm{ht}$. As a result, the spectral weight is gradually redistributed from the original excitonic transition to lower-energy states with increasing CT character. 

A qualitatively similar effect induced by electron- and hole-transfer interactions has previously been predicted by Hestand and Spano within their vibronic FE--CT model.\cite{hestand_jcp_15} Despite the substantial differences between the two theoretical frameworks, both approaches describe the same physical mechanism by which charge-transfer interactions lead to redistribution of absorption intensity between excitonic and CT-dominated states.

\subsubsection{Structure of the CT-Driven Electronic Spectrum}

We now examine the electronic absorption spectrum $S_0(\hbar\omega)$ in detail, focusing on the influence of charge-transfer interactions on both the global spectral profile and the individual absorption bands. We consider the general case where both excitonic states of the dimer are optically active and coupled to the CT manifold, giving rise to four spectral components associated with the adiabatic excited states. As illustrated by the energy-gap diagram in Fig.~\ref{fig:spectra}B, these components remain energetically separated even in the presence of pronounced CT-induced broadening. This separation allows one to characterize not only the spectrum as a whole, but also the individual absorption bands in terms of their integrated intensities, spectral positions, and effective widths.

To quantify the global properties of the electronic spectrum $S_0(E)$, we introduce the spectral centroid $\bar{E}$ and the effective width $\Delta$, defined through the first and second central moments of the absorption profile,
\begin{equation}\label{spectrum_moments_def}
   \bar{E} \equiv \int E\, S_0(E)\, dE, \qquad
   \Delta^2 \equiv \int (E - \bar{E})^2\, S_0(E)\, dE,
\end{equation}
where the integration is performed over the entire energy range. The spectrum $S_0(E)$ given by Eq.~\eqref{spectrum_def} satisfies the normalization condition
\begin{equation*}
   \int S_0(E)\, dE = 1.
\end{equation*}

Within numerical accuracy, the spectral centroid $\bar{E}$ is found to be invariant upon variation of the CT-related electronic couplings ($V_\mathrm{et}$, $V_\mathrm{ht}$, and $V_\mathrm{ceht}$) and the parameters characterizing charge separation in the dimer ($\Delta E_\mathrm{cs}$ and $\lambda_\mathrm{cs}$). Thus, neither exciton--CT mixing nor solvent-induced stabilization of the zwitterionic states changes the mean transition energy. In contrast, $\bar{E}$ strongly depends on the excitonic coupling $V_\mathrm{ext}$ and on the relative orientation of the local TDMs, characterized by $\theta$. This dependence can be understood directly within the pure excitonic model. Substituting the doublet spectrum $S_0^{(\mathrm{FE})}$ from Eq.~\eqref{spectrum_2delta} into Eq.~\eqref{spectrum_moments_def} gives
\begin{equation}\label{bar_E_2exciton}
   \bar{E} = E_\mathrm{ex} + V_\mathrm{ext}\cos\theta.
\end{equation}
Since the CT interactions do not affect the first spectral moment, Eq.~\eqref{bar_E_2exciton} remains valid for nonzero values of $V_\mathrm{et}$, $V_\mathrm{ht}$, and $V_\mathrm{ceht}$. Thus, while CT interactions can substantially modify the distribution of oscillator strength among the individual absorption bands, as discussed below, they do not shift the overall spectral centroid.

\begin{figure}[ht] 
  \centering 
  \includegraphics[width=10 cm]{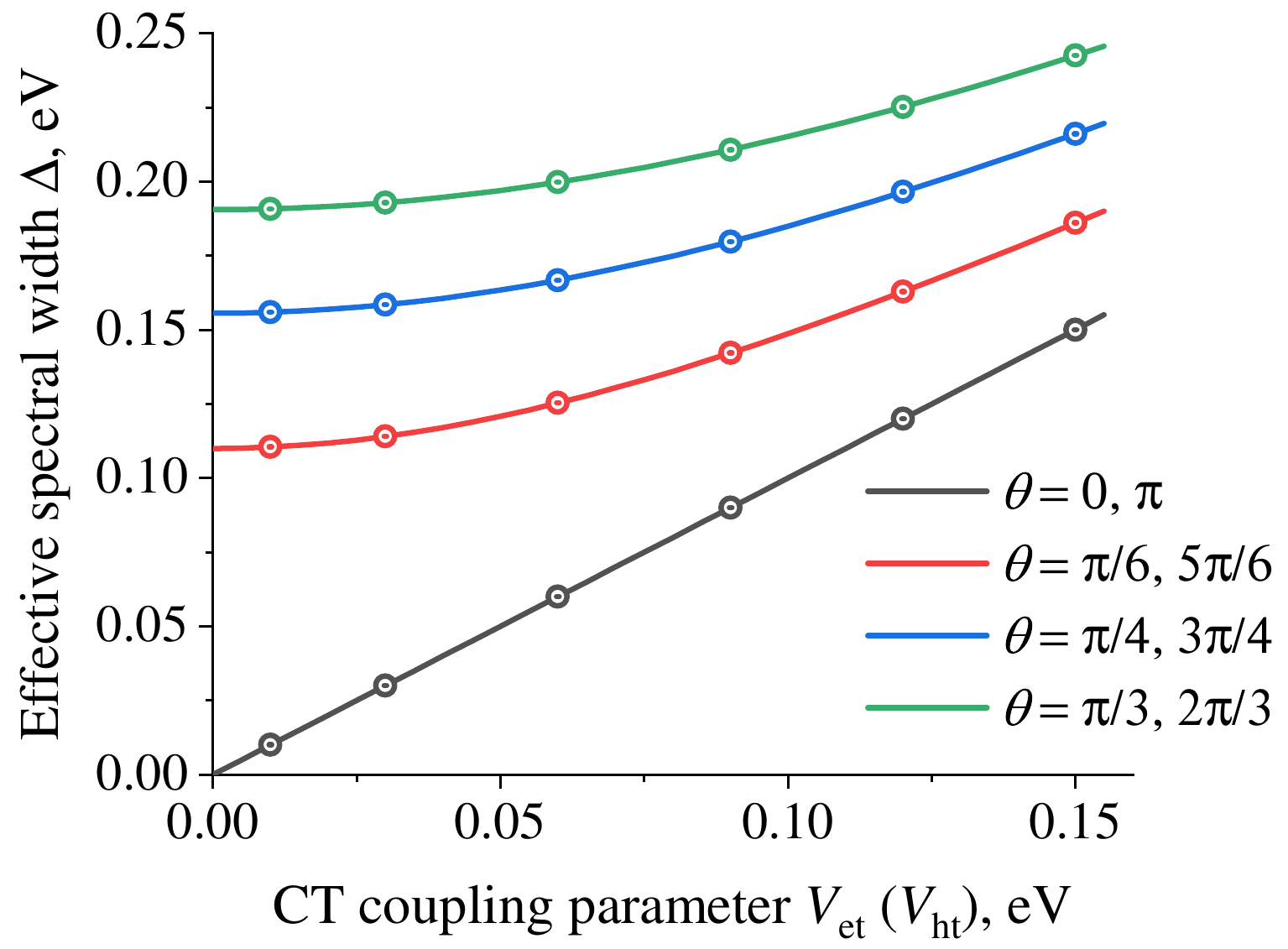} 
    \caption{Effective spectral width $\Delta$ of the electronic absorption spectrum $S_0$ as a function of the CT coupling parameter $V_\mathrm{et}$ ($V_\mathrm{ht}$) for different values of the angle $\theta$. Symbols show numerical results obtained from Eqs.~\eqref{spectrum_moments_def} and \eqref{spectrum_def}, while solid curves represent the analytical approximation given by Eq.~\eqref{Delta_approximation}. In the $\Delta(V_\mathrm{et})$ calculations, the parameter $V_\mathrm{ht}$ was set to zero, whereas in the $\Delta(V_\mathrm{ht})$ calculations the parameter $V_\mathrm{et}$ was set to zero. The two dependences coincide within numerical accuracy. Parameters: $V_\mathrm{ceht} = 0$, $V_\mathrm{ext} = 0.22$~eV, $\lambda_\mathrm{cs} = 0.1$~eV, and $\Delta E_\mathrm{cs} = -0.3$~eV.} 
    \label{fig:Delta_broadening} 
\end{figure}

In contrast to the behavior of $\bar{E}$, numerical simulations reveal substantial CT-induced broadening of the electronic spectrum $S_0(E)$. Representative results are shown in Fig.~\ref{fig:Delta_broadening}, where the effective spectral width $\Delta$ is plotted as a function of the CT coupling parameters $V_\mathrm{et}$ and $V_\mathrm{ht}$ for several values of $\theta$. In these calculations, the effect of each coupling parameter was analyzed separately by setting the other one equal to zero. The resulting $\Delta(V_\mathrm{et})$ and $\Delta(V_\mathrm{ht})$ dependences are identical within numerical accuracy, indicating that electron- and hole-transfer interactions contribute equally to CT-induced spectral broadening within the present model.

The effective spectral width $\Delta$ in Fig.~\ref{fig:Delta_broadening} increases markedly with increasing CT coupling strength. This effect is especially pronounced for $\theta = 0$ and $\theta = \pi$ (black curve), where the corresponding $\Delta(V)$ dependences become nearly linear, indicating dominant CT-induced broadening of the electronic spectrum.

An important feature of the CT-induced broadening is that, in the limiting cases $V_\mathrm{ht}=0$, $V_\mathrm{et}\neq 0$ and $V_\mathrm{et}=0$, $V_\mathrm{ht}\neq 0$, the effective spectral width is accurately described by the analytical expression
\begin{equation}\label{Delta_approximation}
    \Delta = \sqrt{\Delta_0^2 + V^2}.
\end{equation}
Here, $V$ denotes the corresponding CT coupling parameter ($V_\mathrm{et}$ or $V_\mathrm{ht}$), while $\Delta_0$ is the effective width of the pure excitonic spectrum $S_0^{(\mathrm{FE})}$. From Eq.~\eqref{spectrum_2delta} one can easily obtain $\Delta_0 = |V_\mathrm{ext}\sin\theta|$. As shown in Fig.~\ref{fig:Delta_broadening}, Eq.~\eqref{Delta_approximation} reproduces the numerical results over the entire range of $\theta$ and $V$ considered in this work.

Equation~\eqref{Delta_approximation} suggests that the spectral broadening originates from two independent physical mechanisms. The first term, $\Delta_0$, arises from excitonic splitting, whereas the CT-related term associated with $V$ reflects exciton--CT mixing. The quadratic form of Eq.~\eqref{Delta_approximation} indicates that these mechanisms contribute to the spectral width in an approximately additive manner.

We now briefly outline the results of numerical analysis of the individual absorption bands; a detailed discussion is provided in Section 3 of the Supporting Information. In the general case, the $S_0(E)$ profile consists of four spectral components associated with optical transitions into the adiabatic excited states $\vert \Psi_k^\mathrm{(a)}(D_\mathrm{m}) \rangle$. The corresponding energy intervals are denoted by $\Omega_k$ ($k = 1,\dots,4$). Their boundaries are determined by the structure of the adiabatic FESs, so that evaluation of $\Omega_k$ requires numerical analysis of the transition-energy profiles $\Delta G_{0k}(D_\mathrm{m})$. As illustrated in Fig.~\ref{fig:spectra}, for sufficiently exergonic charge separation, $-\Delta E_\mathrm{cs} > V_\mathrm{ext}$, bands 1 and 2 correspond to optical transitions into low-energy states with predominantly CT character, whereas bands 3 and 4 are associated primarily with the lower and upper Frenkel excitons, respectively.

For each absorption band, we define the relative intensity $\sigma_k$ as the integral of $S_0(E)$ over the corresponding interval $\Omega_k$. The band center $\epsilon_k$ and the effective width $\delta_k$ are defined through the first and second central moments of the normalized spectral profile within the same interval (Eqs.~(S10)--(S12) in the Supporting Information).

Numerical calculations of the band spectral characteristics reveal several distinct manifestations of exciton--CT mixing in the electronic spectrum of the dimer. In particular, pronounced redistribution of oscillator strength between the individual absorption bands is observed upon variation of the model parameters governing the electronic structure of the coupled FE--CT manifold (see Figs.~S1--S5 in the SI). These parameters include the CT coupling strengths ($V_\mathrm{et}$, $V_\mathrm{ht}$, and $V_\mathrm{ceht}$), the excitonic interaction $V_\mathrm{ext}$, the charge-separation energy gap $\Delta E_\mathrm{cs}$, the solvent reorganization energy $\lambda_\mathrm{cs}$, and the relative orientation of the local TDMs characterized by the angle $\theta$. The strongest redistribution effects are observed for variations of $\Delta E_\mathrm{cs}$ (Fig.~S4) and $\theta$ (Fig.~S5), reflecting substantial reorganization of the adiabatic excited states.

The numerical results also demonstrate state-dependent energetic shifts of the individual absorption bands arising from variations in the FE--CT coupling conditions shown in Figs.~S1--S5. In particular, changes in the dipole--dipole interaction governed by $\theta$ primarily affect the excitonic absorption components (Fig.~S5B), whereas the $V_\mathrm{ceht}$ parameter mainly influences the energetic splitting of the low-energy CT-dominated states (Fig.~S2B). Particularly illustrative are the dependences shown in Fig.~S3B, which demonstrate gradual transformation of the adiabatic states from predominantly excitonic to predominantly CT character upon variation of the charge-separation energy gap $\Delta E_\mathrm{cs}$.

\subsubsection{Vibrational and Environmental Broadening of the Absorption Spectrum}

The electronic absorption spectrum considered above accounts for intermolecular electronic couplings and CT-induced mixing effects, but does not include nuclear relaxation associated with photoexcitation of the dimer. In realistic molecular systems, optical excitation generally induces substantial reorganization of both intramolecular vibrational modes and low-frequency environmental degrees of freedom associated with redistribution of the electronic density in the excited state. These effects introduce additional spectral broadening and vibrational structure and therefore must be incorporated for a realistic description of molecular aggregates.

To extend the model, we separate the nuclear degrees of freedom coupled to electronic excitation into low-frequency modes ($\hbar\Omega \ll k_\mathrm{B}T$), which can be treated classically, and high-frequency modes ($\hbar\Omega \gtrsim k_\mathrm{B}T$), which require a quantum description. The corresponding excitation-induced reorganization free energies are denoted by $\lambda_\mathrm{ex}^{(\mathrm{lf})}$ and $\lambda_\mathrm{ex}^{(\mathrm{hf})}$. We further assume that the solvent reorganization accompanying photoexcitation is independent of the CT-induced solvent polarization introduced above, so that the corresponding nuclear coordinates can be treated as orthogonal. Within this approximation, compact analytical expressions for the vibrationally broadened absorption spectrum can be derived.

Interaction with a single quantum vibrational mode of frequency $\Omega_\mathrm{v}$ and reorganization free energy $\lambda_\mathrm{ex}^{(\mathrm{hf})}$ gives the following expression for the dimer absorption spectrum within the linear-response approximation,
\begin{equation}\label{spectrum_S1_def}
    S_1(E) = \sum\limits_{n=0}^{\infty} F_n \, S_0(E - n\hbar\Omega_\mathrm{v}),
\end{equation}
where $S_0(E)$ is the CT-induced electronic spectrum defined by Eq.~\eqref{spectrum_def}, and $F_n$ is the Franck--Condon factor corresponding to the vibrational transition $\vert 0 \rangle \to \vert n \rangle$
\begin{equation}\label{Fn_def}
    F_n = \frac{S^n e^{-S}}{n!}, \qquad 
    S = \frac{\lambda_\mathrm{ex}^{(\mathrm{hf})}}{\hbar\Omega_\mathrm{v}}.
\end{equation}

We next incorporate broadening induced by low-frequency environmental modes. Assuming that the corresponding nuclear coordinates are uncorrelated with the high-frequency vibrational modes, the low-frequency contribution can be introduced independently through convolution of the vibronically broadened spectrum $S_1(E)$ with the equilibrium distribution of optical energy gaps,
\begin{equation}\label{spectrum_S2_def}
    S_2(E) = \int\limits_{0}^{\infty} S_1(E^\prime)\, F(E - E^\prime)\, dE^\prime,
\end{equation}
where the kernel function $F(E)$ is given by
\begin{equation}\label{F_kernel}
    F(E) = \frac{1}{\sqrt{2\pi\langle E^2 \rangle}}
    \exp\left[
        -\frac{\left(E - 2\lambda_\mathrm{ex}^{(\mathrm{lf})}\right)^2}
        {2\langle E^2 \rangle}
    \right],
\end{equation}
with
\begin{equation*}
    \langle E^2 \rangle =
    2\lambda_\mathrm{ex}^{(\mathrm{lf})}k_\mathrm{B}T.
\end{equation*}

The two broadening mechanisms introduced above produce qualitatively different modifications of the electronic absorption spectrum. Coupling to high-frequency intramolecular vibrations gives rise to vibronic progressions and additional fine structure associated with transitions between quantized vibrational states. In contrast, interaction with low-frequency environmental modes leads primarily to inhomogeneous broadening and smooth Gaussian-like redistribution of spectral intensity. The combined treatment of these effects within the linear-response approximation provides a physically transparent framework for describing realistic absorption spectra of molecular dimers and aggregates. Similar approaches based on Franck--Condon vibronic progressions and Gaussian environmental broadening are widely used in the theory of molecular spectroscopy and excitonic systems. The assumption of independent high- and low-frequency nuclear coordinates corresponds to the standard separation of vibrational and environmental contributions employed in condensed-phase spectroscopy.

\subsection{Excited-State Structure and Absorption Spectrum of a Covalently Linked BPEA Dimer}

As an illustration of the developed theoretical framework, we analyze the excited-state structure and absorption spectrum of the BPEA dimer shown in Fig.~\ref{fig:structures}. The absorption spectra of this bichromophoric compound were previously measured in Ref.~\cite{bae2020} for nonpolar (toluene) and moderately polar (dichloromethane) solvents, revealing pronounced differences from the spectra of the BPEA monomer in the corresponding environments. These experimental observations provide an opportunity to assess the role of interchromophore electronic interactions, exciton--CT mixing, and environmental polarization in shaping the optical response of the dimer. 

We analyze the structure of the absorption spectrum using the theoretical framework developed above in combination with quantum-chemical calculations and available experimental data. The principal spectral components associated with electronic interactions between the BPEA chromophores are identified and characterized. Particular attention is devoted to the spectral signatures of the symmetric and antisymmetric Frenkel exciton states, as well as to the contribution of zwitterionic states to the overall absorption profile.

\subsubsection{Optimized Dimer Geometry}

To determine the equilibrium structure of the BPEA dimer, geometry optimization was performed at the density functional theory (DFT) level using the $\omega$B97X functional with the Grimme D3 dispersion correction\cite{grimme_jcp_10} and the def2-TZVP basis set. This long-range corrected functional provides a balanced description of both locally excited and charge-transfer electronic states, which are central to the present work. Solvent effects for toluene (Tol) and dichloromethane (DCM) were incorporated implicitly using the conductor-like polarizable continuum model (CPCM). The optimized geometries in both solvents were further verified by harmonic vibrational frequency analysis, confirming the absence of imaginary frequencies. All calculations were carried out using the ORCA 6.1.1 software package.\cite{neese_jcp_20, neese_wire_25}

\begin{figure}[ht]
   \centering
   \includegraphics[width=14 cm]{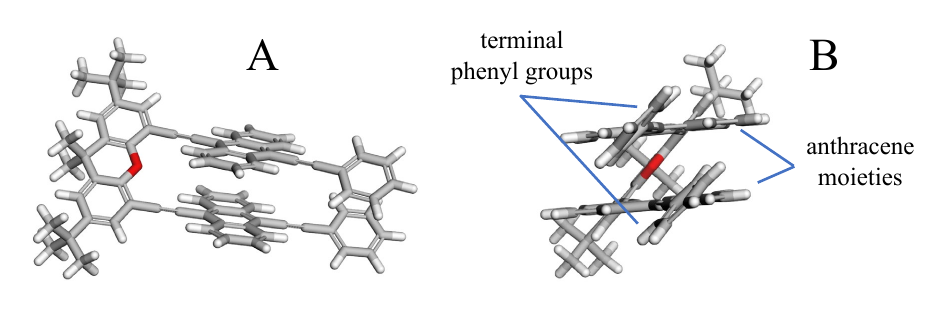}
   \caption{Optimized geometry of the BPEA dimer in DCM. (A) The two BPEA units adopt a folded arrangement driven by $\pi$--$\pi$ interactions between the chromophores. (B) Pronounced torsional rotations of the anthracene and terminal phenyl groups relative to the xanthene linker are observed.}
   \label{fig:dcm_optimized}
\end{figure}

The optimized geometry in dichloromethane (DCM) is shown in Fig.~\ref{fig:dcm_optimized} and reveals a compact folded conformation in which the two anthracene moieties adopt nearly parallel orientations while exhibiting pronounced torsion relative to the xanthene linker ($\delta_\mathrm{x-a} \approx 45^\circ$). The terminal phenyl groups are likewise approximately parallel to one another but display additional torsional displacement relative to the anthracene plane ($\delta_\mathrm{a-p} \approx 40^\circ$), as illustrated in Fig.~\ref{fig:dcm_optimized}B.

The resulting $\pi$-stacked structure, characterized by a face-to-face arrangement of the anthracene and terminal phenyl groups, promotes strong dipole--dipole and CT interactions between the BPEA chromophores, as previously noted in Ref.~\cite{bae2020}. Importantly, however, the long molecular axes of the two BPEA units are not strictly parallel ($\theta \approx 24^\circ$), resulting in a noncollinear orientation of their transition dipole moments. This noncollinearity influences the absorption profile of the dimer by partially lifting the optical forbiddance of the antisymmetric Frenkel exciton state, which is essentially dark in the strictly parallel geometry.

The optimized geometry of the BPEA dimer in toluene is very similar to that obtained in dichloromethane, with only minor variations in the interplanar separations and torsional angles. The key geometric parameters of the DFT-optimized structures in Tol and DCM are summarized in Table~\ref{tab:dimer_geometry}. 

\begin{table}[ht]
    \caption{Structural parameters of the covalently linked BPEA dimer in toluene (Tol) and dichloromethane (DCM). $\theta$ is the angle between the long molecular axes of the two monomeric units; $\gamma_\mathrm{a-a}$ and $\gamma_\mathrm{p-p}$ denote the dihedral angles between the anthracene and phenyl planes of the two BPEA chromophores, respectively; $d_\mathrm{a-a}$ and $d_\mathrm{p-p}$ are the average interplanar separations between the anthracene and terminal phenyl groups; $\delta_\mathrm{x-a}$ and $\delta_\mathrm{a-p}$ represent the torsional angles between the xanthene and anthracene units, and between the anthracene and phenyl groups, respectively. $\kappa$ is the dipole--dipole orientational factor. All angular parameters are given in degrees.}
    
    \centering
    \begin{tabular}{ccccccccc}
    \toprule
    solvent &$\theta$ &$\gamma_\mathrm{a-a}$ & $\gamma_\mathrm{p-p}$ & $d_\mathrm{a-a}$ (\AA) & $d_\mathrm{p-p}$ (\AA) & $\delta_\mathrm{x-a}$ & $\delta_\mathrm{a-p}$ & $\kappa$ \\
    \midrule
    DCM & 23.9 & 0.9 & 11.5 & 3.51 & 3.76 & 44.4 & 39.8 & 0.93 \\
    Tol & 23.8 & 1.2 & 11.6 & 3.48 & 3.71 & 44.5 & 40.6 & 0.93 \\
    \bottomrule
    \end{tabular}
    
    \label{tab:dimer_geometry}
\end{table}

The anthracene fragments remain nearly parallel ($\gamma_\mathrm{a-a}\approx1^\circ$), whereas the terminal phenyl groups exhibit a somewhat larger mutual inclination ($\gamma_\mathrm{p-p}\approx12^\circ$) owing to the torsional flexibility of the peripheral units. The average interplanar separations ($d_\mathrm{a-a}\approx3.5$ Å and $d_\mathrm{p-p}\approx3.7$ Å) are characteristic of efficient $\pi$-stacking interactions. The corresponding dipole--dipole orientational factor ($\kappa\approx0.93$) indicates that the folded geometry remains favorable for strong excitonic interactions despite the finite angle between the molecular axes.

Next, we investigate the excited electronic states of the dimer in the equilibrium solvent environment using time-dependent density functional theory (TDDFT). The resulting excitation energies and transition dipole moments are analyzed within the Kasha vector model, in which the Frenkel exciton states are represented as symmetric and antisymmetric linear combinations of local monomer excitations. The corresponding monomer properties are used as reference quantities in this analysis.

\subsubsection{Excited-State Energies and Optical Properties of the Dimer}

The excited electronic states of the BPEA dimer were computed using TDDFT at the same level of theory and with the same solvent model as employed for the ground-state geometry optimization. To facilitate interpretation of these states within the Frenkel exciton framework, additional TDDFT calculations were performed for the isolated BPEA monomer in both solvents.

\begin{figure}[ht]
   \centering
   \includegraphics[width=15 cm]{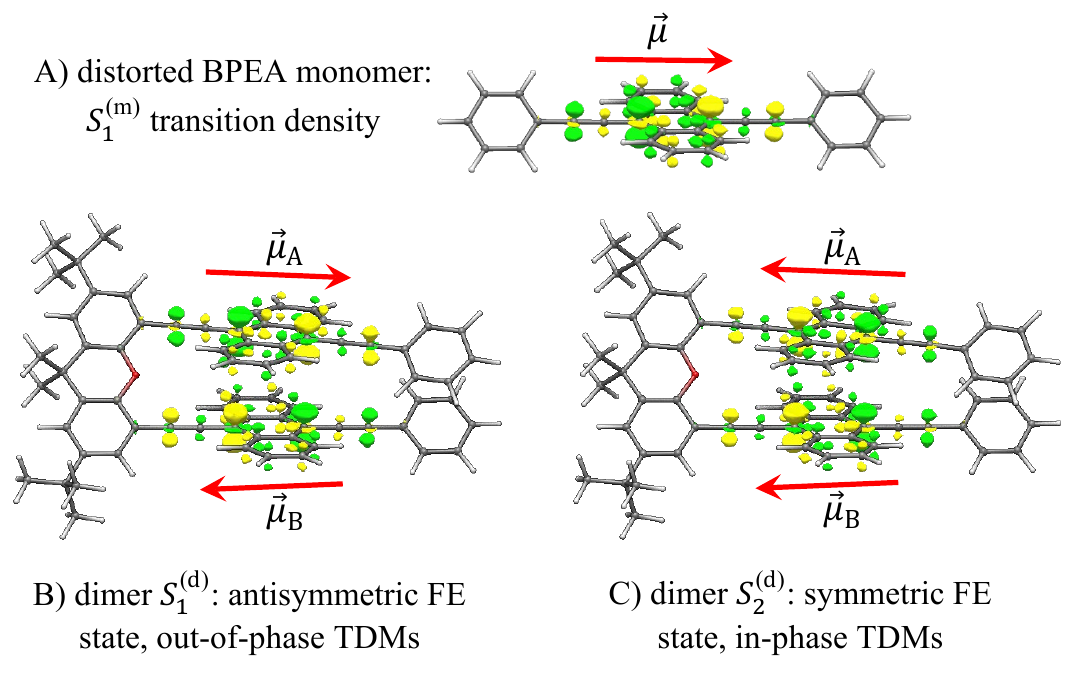}
   \caption{Electronic transition densities of (A) the lowest singlet excited state of the distorted BPEA monomer and (B, C) the two lowest excitonic states of the covalent BPEA dimer. Panel B corresponds to the antisymmetric (dark) state $S_1^\mathrm{(d)}$, characterized by an out-of-phase combination of monomer-like transition densities on the two chromophore units. Panel C shows the symmetric (bright) state $S_2^\mathrm{(d)}$, in which the monomer transition densities appear in phase.}
   \label{fig:dimer_TDs}
\end{figure}

The monomer geometries were generated by fragmenting the optimized dimer structure into two individual chromophoric units while retaining the distortions induced by dimer formation. In this procedure, the carbon framework of each chromophore was preserved, whereas the bonds to the xanthene linker were cleaved and saturated with hydrogen atoms. The added hydrogen atoms were subsequently optimized in the presence of the solvent with all carbon positions held fixed. The resulting distorted monomer structure (see Fig.~\ref{fig:dimer_TDs}A) was then used in the TDDFT analysis of the excited states and the corresponding transition dipole moments.

For the distorted monomer geometry, the lowest optically active singlet excitation $S^{(\mathrm{m})}_1$ is located at $E^{(\mathrm{m})}_1 = 2.96$ eV in DCM and $E^{(\mathrm{m})}_1 = 2.95$ eV in Tol. The associated transition dipole moments are $\mu^{(\mathrm{m})}_1 = 8.71$ D in DCM and $\mu^{(\mathrm{m})}_1 = 8.82$ D in Tol, with the $\bm{\mu}^{(\mathrm{m})}_1$ vector oriented essentially along the long molecular axis of the chromophore.

\begin{table}[ht]
    \caption{TDDFT excitation energies $E_k^\mathrm{(d)}$, transition dipole magnitudes $\mu_k^\mathrm{(d)}$, and corresponding unit vectors $\bm{n}_k^\mathrm{(d)} = \bm{\mu}_k^\mathrm{(d)}/|\bm{\mu}_k^\mathrm{(d)}|$ for the four lowest singlet states ($S_1^\mathrm{(d)} - S_4^\mathrm{(d)}$) of the BPEA dimer in dichloromethane and toluene. The components of the $\bm{n}_k^\mathrm{(d)}$ vectors are given in the molecular coordinate system, where the $X$ axis is aligned with the long molecular axis of the bichromophore and the $Y$ axis lies in the xanthene plane. Energies are given in eV and dipole moments in Debye.}
    
    \centering
    \begin{tabular}{c cc cc cc}
    \toprule
    State 
    & \multicolumn{2}{c}{$E_k^\mathrm{(d)}$ (eV)} 
    & \multicolumn{2}{c}{$\mu_k^\mathrm{(d)}$ (D)} 
    & \multicolumn{2}{c}{$\bm{n}_k^\mathrm{(d)}$} \\
    
    \cmidrule(lr){2-3}
    \cmidrule(lr){4-5}
    \cmidrule(lr){6-7}
    
    & DCM & Tol 
    & DCM & Tol 
    & DCM & Tol \\
    
    \midrule
    
    $S_1^\mathrm{(d)}$ & 2.86 & 2.86 & 1.90 & 1.93 & 
    (0.06, -0.61, 0.79) & (0.06, -0.63, 0.78) \\
    
    $S_2^\mathrm{(d)}$ & 2.98 & 2.98 & 11.33 & 11.49 & 
    (1.00, -0.02, 0.00) & (1.00, -0.02, 0.00) \\
    
    $S_3^\mathrm{(d)}$ & 3.60 & 3.60 & 1.68 & 1.57 & 
    (0.89, -0.39, -0.25) & (0.80, -0.51, -0.31) \\
    
    $S_4^\mathrm{(d)}$ & 3.61 & 3.61 & 1.76 & 1.98 & 
    (0.91, -0.41, -0.02) & (0.95, -0.31, -0.03) \\
    
    \bottomrule
    \end{tabular}
    
    \label{tab:excited_states}
\end{table}

The TDDFT results for the covalently linked dimer are summarized in Table~\ref{tab:excited_states}, which reports the excitation energies and transition dipole moments of the four lowest singlet excited states $S_k^\mathrm{(d)}$ ($k = 1\text{--}4$) in DCM and Tol. The pronounced difference in the transition dipole magnitudes of the two lowest states ($\mu_1^\mathrm{(d)} = 1.90$~D and $\mu_2^\mathrm{(d)} = 11.33$~D in DCM) indicates that $S_1^\mathrm{(d)}$ and $S_2^\mathrm{(d)}$ form an excitonic doublet with the level ordering characteristic of an H-type bichromophore. Within the Kasha model, these states are represented as symmetric and antisymmetric linear combinations of local monomer excitations,
\begin{equation}\label{FE_wfs}
   \vert S_1^\mathrm{(d)} \rangle = \frac{1}{\sqrt{2}}\left( \vert M^*_1 M_2 \rangle - \vert M_1 M^*_2\rangle \right), \qquad
   \vert S_2^\mathrm{(d)} \rangle = \frac{1}{\sqrt{2}}\left( \vert M^*_1 M_2 \rangle + \vert M_1 M^*_2\rangle \right).
\end{equation}

This interpretation is further supported by the TDDFT transition densities of the $S_1^\mathrm{(d)}$ and $S_2^\mathrm{(d)}$ states, $\rho_1^{\mathrm{(d)}}(\bm{r})$ and $\rho_2^{\mathrm{(d)}}(\bm{r})$, shown in Fig.~\ref{fig:dimer_TDs}. The transition density of the lowest excited state, $\rho_1^{\mathrm{(d)}}(\bm{r})$, exhibits an antisymmetric distribution over the two chromophore units, consistent with an out-of-phase arrangement of their local TDMs $\bm{\mu}_\mathrm{A}$ and $\bm{\mu}_\mathrm{B}$ (Fig.~\ref{fig:dimer_TDs}B). In contrast, $\rho_2^{\mathrm{(d)}}(\bm{r})$ displays a symmetric distribution indicative of an in-phase alignment of $\bm{\mu}_\mathrm{A}$ and $\bm{\mu}_\mathrm{B}$ (Fig.~\ref{fig:dimer_TDs}C). 

It is also instructive to compare the transition density of the distorted BPEA monomer ($\rho_1^{\mathrm{(m)}}(\bm{r})$, see Fig.~\ref{fig:dimer_TDs}A) with the corresponding transition densities of the lowest dimer states. The spatial profiles of $\rho_1^{\mathrm{(d)}}(\bm{r})$ and $\rho_2^{\mathrm{(d)}}(\bm{r})$ closely resemble linear combinations of two monomer transition densities localized on the individual chromophore units. In the $S_2^\mathrm{(d)}$ state (Fig.~\ref{fig:dimer_TDs}C), the monomer-like transition densities appear in phase on the two BPEA units, producing an enhanced symmetric transition density distribution. In contrast, the $S_1^\mathrm{(d)}$ state (Fig.~\ref{fig:dimer_TDs}B) exhibits an out-of-phase combination of the same monomer transition densities, leading to the antisymmetric spatial pattern. This clear correspondence between the monomer and dimer transition densities provides further evidence for the predominantly Frenkel exciton character of the two lowest excited states with only negligible charge-transfer admixture.

\subsubsection{Vector-Model Analysis of the Excitonic Doublet}

Eq.~\ref{FE_wfs} provides the basis for a quantitative analysis of the TDDFT results for the BPEA dimer within the Kasha vector model. In this simplified picture, the transition dipole moments of the symmetric and antisymmetric exciton states arise from vector superposition of the local transition dipoles ($\bm{\mu}_\mathrm{A}$ and $\bm{\mu}_\mathrm{B}$) associated with the individual chromophores,
\begin{equation}\label{FE_tdms}
   \bm{\mu}_- = \frac{1}{\sqrt{2}}\left( \bm{\mu}_\mathrm{A} - \bm{\mu}_\mathrm{B} \right), \qquad
   \bm{\mu}_+ = \frac{1}{\sqrt{2}}\left( \bm{\mu}_\mathrm{A} + \bm{\mu}_\mathrm{B} \right).
\end{equation}
Assuming equal magnitudes $\mu_0 = |\bm{\mu}_\mathrm{A}| = |\bm{\mu}_\mathrm{B}|$ and an angle $\theta$ between the two local dipoles, one obtains the following expressions for the transition dipole magnitudes,
\begin{equation}\label{FE_magnitudes}
   \mu_- \equiv \left| \bm{\mu}_- \right| = \sqrt{2}\,\mu_0 \left| \sin\frac{\theta}{2} \right|, \qquad
   \mu_+ \equiv \left| \bm{\mu}_+ \right| = \sqrt{2}\,\mu_0 \left| \cos\frac{\theta}{2} \right|.
\end{equation}

Using the magnitude of the monomer TDM in dichloromethane ($\mu_0 = 8.71$~D) and the angle $\theta = 23.9^\circ$ between the local dipoles in the DCM-optimized dimer geometry, Eqs.~\eqref{FE_magnitudes} yield
\[
\mu_- \approx 2.56~\mathrm{D}, 
\qquad
\mu_+ \approx 12.1~\mathrm{D}.
\]
These values are in reasonable agreement with the TDDFT results $\mu_1^\mathrm{(d)} = 1.90$~D and $\mu_2^\mathrm{(d)} = 11.3$~D, indicating that the simple vector superposition model captures the essential features of the two lowest excited states. The remaining discrepancies likely originate from weak asymmetry between the chromophores and from limitations of the point-dipole approximation at short intermolecular separations.

\begin{figure}[ht]
   \centering
   \includegraphics[width=14 cm]{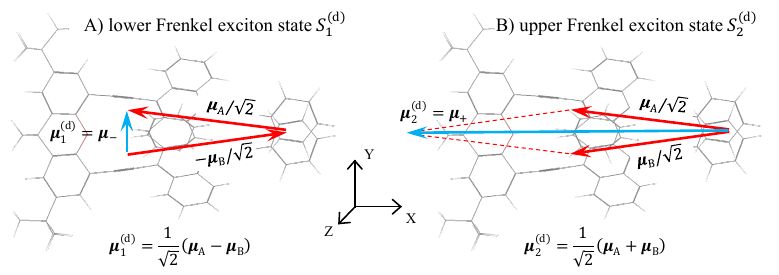}
   \caption{Simple vector model for the FE states of the BPEA dimer. The local transition dipole moments $\bm{\mu}_\mathrm{A}/\sqrt{2}$ and $\bm{\mu}_\mathrm{B}/\sqrt{2}$ are shown together with their (A) vector difference $\bm{\mu}_-$ and (B) vector sum $\bm{\mu}_+$, corresponding to the lower and upper excitonic states $S_1^\mathrm{(d)}$ and $S_2^\mathrm{(d)}$, respectively.}
   \label{fig:vector_model}
\end{figure}

A geometric illustration of the vector model is presented in Fig.~\ref{fig:vector_model}, showing the local TDMs of the two BPEA units together with their vector sum and difference, corresponding to the symmetric and antisymmetric FE states. Constructive interference between $\bm{\mu}_\mathrm{A}$ and $\bm{\mu}_\mathrm{B}$ (panel B) produces an enhanced transition dipole $\bm{\mu}_+$ for the upper (bright) state. In contrast, the vector difference (panel A) reflects destructive interference between the local TDMs, resulting in strongly reduced optical activity of the lower excitonic state. The finite angle $\theta$ between the two dipoles prevents complete cancellation, thereby accounting for the residual intensity of the nominally dark antisymmetric state.

The vector model also provides a straightforward interpretation of the orientation of the excitonic TDMs $\bm{\mu}_1^\mathrm{(d)}$ and $\bm{\mu}_2^\mathrm{(d)}$ in the molecular coordinate system of the BPEA dimer. As illustrated in Fig.~\ref{fig:vector_model}, the vector sum $\bm{\mu}_+$ is directed predominantly along the long molecular axis ($X$) of the bichromophore, whereas the vector difference $\bm{\mu}_-$ retains a substantial transverse component and is therefore oriented largely perpendicular to the $X$ axis. This geometric picture is consistent with the TDDFT results summarized in Table~\ref{tab:excited_states}, which show that $\bm{\mu}_2^\mathrm{(d)}$ is aligned mainly along $X$, while $\bm{\mu}_1^\mathrm{(d)}$ is oriented predominantly in the transverse direction.

\subsubsection{Analysis and Decomposition of the Absorption Spectrum in Dichloromethane}

In this subsection, we analyze the absorption spectrum of the BPEA dimer in DCM with the aim of identifying the contributions of Frenkel exciton and zwitterionic (charge-transfer) states and elucidating the roles of excitation-induced intramolecular vibrations and solvent reorganization in shaping the spectral profile. The analysis is carried out in two steps. First, the experimental spectrum is fitted using a vibronic-exciton model that accounts for electronic interactions between the chromophores, coupling to high-frequency (hf) intramolecular vibrational modes, and low-frequency (lf) solvent reorganization. In this procedure, the key electronic parameters are taken from quantum-chemical calculations, whereas the remaining parameters, primarily those describing vibronic and environmental effects, are determined by fitting to the experimental data. In the second step, the resulting parameter set is used to decompose the spectrum by selectively including or excluding specific interactions, thereby isolating the effects of electronic coupling, vibronic structure, and solvent-induced broadening. This approach enables interpretation of the observed spectral features in terms of excitonic splitting, vibronic progression, and mixing with CT states in a polar environment.

For the covalently linked BPEA dimer in dichloromethane, several electronic parameters entering the model are determined independently from quantum-chemical calculations. The excitonic coupling, $V_\mathrm{ext} = 0.0625$~eV, is obtained from the energy splitting between the two lowest singlet excited states of the dimer in DCM, $S_1^{\mathrm{(d)}}$ and $S_2^{\mathrm{(d)}}$ (Table~\ref{tab:excited_states}). The electron- and hole-transfer couplings, $V_\mathrm{et} = 0.0596$~eV and $V_\mathrm{ht} = -0.0315$~eV, together with the charge-separation energy, $\Delta E_\mathrm{cs} = 0.132$~eV, are taken from previous quantum-chemical calculations reported in Ref.~\cite{bae2020}. The angle between the local TDMs, $\bm{\mu}_\mathrm{A}$ and $\bm{\mu}_\mathrm{B}$, is $\theta = 23.9^\circ$. The concerted electron--hole transfer interaction is neglected in the present model by setting $V_\mathrm{ceht} = 0$.

To assess the sensitivity of the dimer geometry to the computational level, we additionally optimized the BPEA dimer at the B3LYP-D3/6-31G* level used in Ref.~\cite{bae2020}. The resulting structure retains the same compact folded arrangement as the $\omega$B97X-D3/def2-TZVP structure used in the present work, with the anthracene and phenyl fragments of the two monomers oriented nearly parallel to each other. The intermolecular distances between the corresponding aromatic planes are also similar, whereas moderate differences are found in the torsional angles and in the angle between the local TDMs. Thus, the two computational approaches yield qualitatively similar dimer geometries, supporting the use of the parameters from Ref.~\cite{bae2020} in the present analysis, while some quantitative uncertainty associated with the different computational levels cannot be excluded.

The remaining parameters primarily characterize coupling of the electronic excitations to intramolecular vibrations and to the solvent environment. These adjustable quantities include the excitation-related intramolecular (high-frequency) reorganization energy $\lambda_\mathrm{ex}^\mathrm{(hf)}$, the solvent (low-frequency) reorganization energy $\lambda_\mathrm{ex}^\mathrm{(lf)}$, the frequency of the effective intramolecular vibrational mode $\Omega_\mathrm{v}$, and the excitation energy $E_\mathrm{ex}$. Their values are determined by fitting the calculated total absorption spectrum $S_2(\hbar\omega)$ [Eq.~\eqref{spectrum_S2_def}] to the experimental absorption profile of the BPEA dimer in DCM.\cite{bae2020}

\begin{figure}[ht]
   \centering
   \includegraphics[width=14 cm]{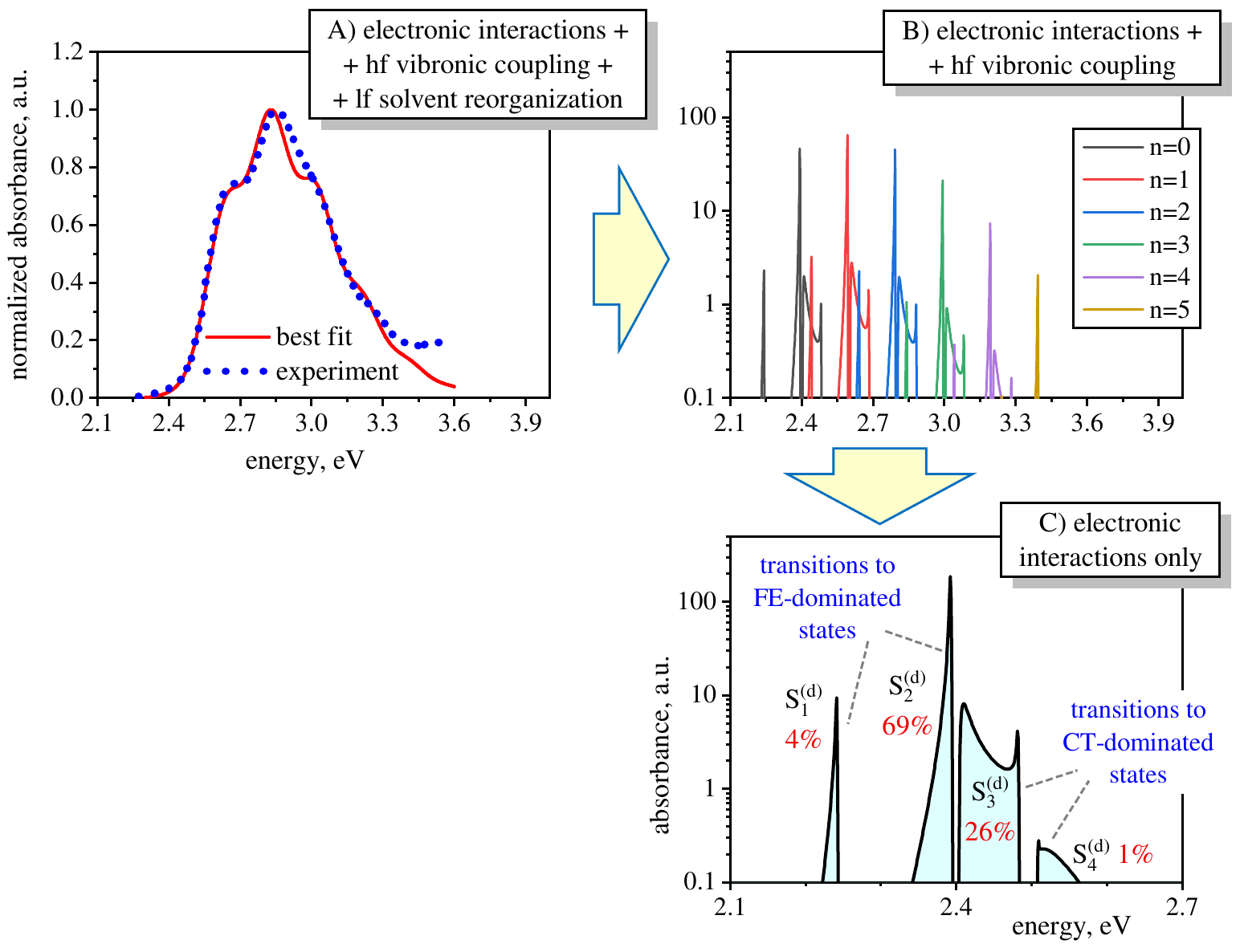}
   \caption{Spectral analysis of the absorption band of the BPEA dimer in dichloromethane. (A) Experimental (blue) and fitted (red) absorption spectra including electronic interactions, vibronic coupling, and solvent reorganization. (B) Vibronically broadened spectrum $S_1(\hbar\omega)$ obtained with high-frequency intramolecular vibrations only, without low-frequency solvent broadening [Eq.~\eqref{spectrum_S1_def}], showing the resolved vibronic progression (logarithmic Y-scale). (C) Electronic spectrum $S_0(\hbar\omega)$ [Eq.~\eqref{spectrum_def}] containing only excitonic and CT interactions (logarithmic Y-scale). The percentages indicate the relative integrated intensities $\sigma_k$ of the four electronic absorption bands.}
   \label{fig:fitted_spectra}
\end{figure}

The fitting procedure was performed over the energy interval from 2.4 to 3.3~eV, encompassing the experimentally observed absorption band associated with the excitonic transition. The optimization involved minimization of the least-squares deviation between the calculated and experimental spectra. The resulting fit is shown in Fig.~\ref{fig:fitted_spectra}A. Owing to the pronounced sensitivity of the spectral profile to variations in the adjustable parameters, the fitting procedure yields well-defined optimal values. The resulting parameters are
\begin{equation}
   \lambda_\mathrm{ex}^\mathrm{(lf)} = 0.11\ \mathrm{eV}, \qquad
   \lambda_\mathrm{ex}^\mathrm{(hf)} = 0.28\ \mathrm{eV}, \qquad
   \hbar \Omega_\mathrm{v} = 0.19\ \mathrm{eV}, \qquad
   E_\mathrm{ex} = 2.34\ \mathrm{eV}.
\end{equation}
As shown in Fig.~\ref{fig:fitted_spectra}A, these parameters successfully reproduce both the vibronic progression and the overall spectral envelope of the experimental absorption band.

To elucidate the roles of different physical contributions to the spectral profile, two additional calculations are presented in Fig.~\ref{fig:fitted_spectra}B,C. Figure~\ref{fig:fitted_spectra}B shows the spectrum $S_1(\hbar\omega)$ obtained when only high-frequency vibronic coupling is included, while low-frequency solvent reorganization is neglected according to Eq.~\eqref{spectrum_S1_def}. In this case, the vibronic progression becomes clearly resolved, revealing the vibrational structure of the absorption band. For clarity, the spectrum is plotted on a logarithmic scale, which emphasizes the weaker higher-order vibronic transitions.

Furthermore, Fig.~\ref{fig:fitted_spectra}C shows the purely electronic absorption profile $S_0(\hbar\omega)$ calculated using Eq.~\eqref{spectrum_def} without high- and low-frequency nuclear reorganization effects. In this case, the spectrum reflects only the electronic interactions between the two chromophores, including the excitonic and CT couplings introduced in Section~\ref{sec:methods}. Four distinct absorption bands corresponding to optical transitions into the adiabatic excited states $S_1^{\mathrm{(a)}}$ -- $S_4^{\mathrm{(a)}}$ can be identified in the profile $S_0(\hbar\omega)$. In contrast to the strongly exergonic regime considered in Section~\ref{sec:spectra}, the BPEA dimer in DCM is characterized by a positive charge-separation energy, $\Delta E_\mathrm{cs} = 0.132$~eV, resulting in CT states located above the excitonic manifold. Consequently, the two lower-energy bands correspond predominantly to the excitonic doublet, whereas the higher-energy absorption components are associated with states possessing substantial CT character. The relative integrated intensities $\sigma_k$ of these electronic bands are indicated in the figure. Taken together, Fig.~\ref{fig:fitted_spectra}A--C illustrates how the full absorption spectrum emerges from the combined effects of interchromophoric electronic interactions, vibronic coupling, and solvent reorganization, with each contribution progressively shaping the spectral envelope and intensity distribution.

The results obtained also provide insight into the physical mechanisms underlying the observed spectral broadening. In particular, the relatively large reorganization energies, especially $\lambda_\mathrm{ex}^\mathrm{(lf)}$, indicate substantial coupling of the electronic excitation to both intramolecular vibrations and the surrounding solvent environment. This behavior is observed despite the absence of pronounced charge separation upon photoexcitation of the BPEA chromophore. In contrast to typical donor--acceptor systems, where excitation is accompanied by strong intramolecular charge transfer, photoexcitation of BPEA primarily involves redistribution of electronic density while preserving an approximately symmetric charge distribution.\cite{SIPLIVYSynMet2026} The pronounced solvent response therefore reflects the sensitivity of the surrounding medium to local changes in the electronic density, emphasizing the importance of short-range electrostatic interactions in the optical excitation of symmetric $\pi$-conjugated chromophores and their aggregates. In addition, partial admixture of CT configurations into the excitonic states may further enhance coupling to the polar environment and contribute to the observed spectral broadening. This interpretation is consistent with the analysis of the purely electronic spectrum in Fig.~\ref{fig:fitted_spectra}C, where the two lower-energy states are predominantly excitonic but exhibit non-negligible CT admixture, whereas the higher-energy states possess primarily CT character. 

Despite the substantial contribution of solvent reorganization to the overall spectral broadening, its influence on the electronic structure of the BPEA dimer remains relatively weak. Consequently, the absorption spectra calculated for solvents of different polarity are expected to differ only slightly, consistent with the experimental spectra reported for toluene and dichloromethane.\cite{bae2020} This behavior reflects the fact that solvent polarization primarily broadens the ensemble of transition energies, whereas the underlying excitonic and charge-transfer electronic structure remains essentially unchanged.

\section{Conclusions}

In this paper, we developed a theoretical framework for the analysis of electronic absorption spectra of molecular dimers exhibiting simultaneous Frenkel exciton (FE) and charge-transfer (CT) interactions. The model explicitly incorporates excitonic coupling, electron- and hole-transfer interactions, concerted electron--hole transfer, and solvent-induced stabilization of zwitterionic configurations within a unified adiabatic-state formalism. Analytical expressions were derived for the electronic absorption spectrum and its spectral moments, allowing direct characterization of the spectral centroid, overall spectral width, and individual absorption bands associated with the adiabatic excited states of the coupled FE--CT manifold.

The analysis reveals several characteristic spectral manifestations of exciton--CT mixing. In particular, CT interactions were found to produce substantial redistribution of oscillator strength between the absorption bands and pronounced broadening of the overall spectral profile. Numerical calculations demonstrate that the dominant mechanism of CT-induced broadening originates primarily from additional energetic splitting between spectral components rather than from broadening of the individual bands themselves. At the same time, the first spectral moment remains insensitive to the CT interaction parameters and is governed predominantly by dipole--dipole coupling within the excitonic manifold. In the limiting cases where either electron- or hole-transfer coupling dominates, the overall spectral width is accurately described by a simple analytical expression combining excitonic and CT contributions in a quadratic form, indicating that these broadening mechanisms act in an approximately independent manner.

To enable realistic modeling of experimental spectra, the electronic model was further extended to include coupling to high-frequency intramolecular vibrations and low-frequency environmental degrees of freedom. Within the assumption of independent nuclear coordinates associated with the two mechanisms, compact expressions were obtained for the vibronically and environmentally broadened absorption spectra. This formulation provides a practical framework for decomposition of experimental spectral profiles into electronic, vibronic, and solvent-induced contributions.

The developed approach was applied to the analysis of the absorption spectrum of a covalently linked BPEA dimer in dichloromethane. Quantum-chemical calculations reveal that the two lowest excited states form an excitonic doublet characteristic of an H-type bichromophore, while higher-energy states possess substantial CT character. The experimentally observed absorption spectrum was successfully reproduced using a combined vibronic--exciton model including excitonic coupling, CT interactions, intramolecular vibrational progression, and solvent reorganization. Decomposition of the fitted spectrum demonstrates how the experimentally observed spectral envelope emerges progressively from purely electronic interactions, vibronic coupling, and low-frequency solvent broadening.

The analysis further shows that substantial solvent-induced broadening may arise even in symmetric $\pi$-conjugated chromophores lacking pronounced photoinduced charge separation. In the BPEA dimer, the strong solvent response is attributed to redistribution of electronic density upon excitation and to partial admixture of CT configurations into the excitonic states. These results emphasize the importance of local electrostatic interactions and exciton--CT mixing in determining the optical response of molecular aggregates in polar environments.

The present treatment is formulated within the adiabatic approximation, in which the purely electronic Hamiltonian is diagonalized prior to incorporation of vibronic effects through Franck--Condon progressions associated with the resulting adiabatic states. Consequently, explicit non-adiabatic vibronic coupling between excitonic and charge-transfer states is not included. Such coupling may become significant in molecular systems where the energetic separation between FE and CT states approaches the energy of strongly coupled intramolecular vibrations. Experimental evidence for the role of vibronic FE–CT interactions in molecular dimers has recently been reported by Lin et al.\cite{lin_nc_22} Explicit treatment of these interactions within a fully vibronic Hamiltonian, where electronic and vibrational degrees of freedom are treated simultaneously, has proven successful in describing FE–CT mixing in molecular aggregates and donor–acceptor systems.\cite{hestand_jcp_15} Extension of the present analytical framework to include explicit FE--CT vibronic coupling therefore represents an interesting direction for future work.

Overall, the proposed theoretical framework establishes a physically transparent approach for analysis and decomposition of complex absorption spectra in multichromophoric systems with coupled excitonic and CT states. The methodology developed here should be applicable to a broad class of $\pi$-stacked molecular aggregates, bichromophoric systems, and organic electronic materials where electronic delocalization and charge-transfer interactions jointly govern the excited-state properties. Since the formulation is based on a general diabatic representation of coupled excitonic and charge-transfer configurations, the proposed methodology is applicable to molecular systems with different excited-state ordering and electronic coupling regimes.

\section*{Acknowledgements}

This study was supported by the Russian Science Foundation (Project No. 22-13-00180-P, https://rscf.ru/en/project/22-13-00180/).

\section*{Supporting information}

The following files are available free of charge.
\begin{itemize}
  \item SI.pdf: Solvent polarization coordinate and charge-dissymmetry operator; Optical activity of mixed FE-CT configurations in dimers; Spectral characteristics of electronic absorption bands; Comparison of the BPEA dimer geometries obtained at different computational levels; Computational protocol for absorption spectra
  \item optimized\_wb97x-d3\_dcm.xyz: Optimized geometry of the BPEA dimer in dichloromethane calculated at the $\omega$B97X-D3/def2-TZVP level
  \item optimized\_wb97x-d3\_tol.xyz: Optimized geometry of the BPEA dimer in toluene calculated at the $\omega$B97X-D3/def2-TZVP level
  \item optimized\_b3lyp-d3\_dcm.xyz: Optimized geometry of the BPEA dimer in dichloromethane calculated at the B3LYP-D3/6-31G* level 
\end{itemize}

\bibliography{biblio_dimers}

\newpage

\begin{center}
    TOC graphic
\end{center}

\begin{figure}[ht]
   \centering
   \includegraphics[width=15 cm]{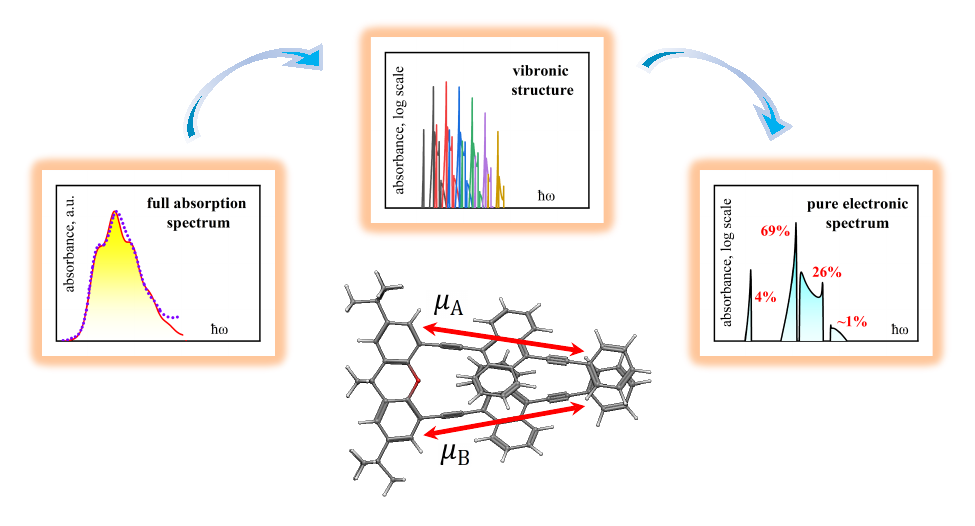}
\end{figure} 

\end{document}